\documentclass[12pt]{article}
\def\Slash#1{#1\kern-0.55em\raise.05ex\hbox{/}}
\usepackage{amsmath,amssymb,amsfonts,amscd,bbm,cancel,graphicx,hyperref}
\begin{document}


\thispagestyle{empty}
\renewcommand{\thefootnote}{\fnsymbol{footnote}}
\setcounter{footnote}{1}

\vspace*{-1.cm}
\begin{flushright}
OSU-HEP-06-5 \\
TUM-HEP-644/06
\end{flushright}
\vspace*{1.8cm}

\centerline{\Large\bf Discretized Gravity in 6D Warped Space}

\vspace*{18mm}

\centerline{\large\bf
Florian Bauer$^a$\footnote{E-mail: \texttt{fbauer@ph.tum.de}},
Tomas H\"allgren$^b$\footnote{E-mail: \texttt{tomashal@kth.se}},
and Gerhart Seidl$^c$\footnote{E-mail: \texttt{seidl@physik.uni-wuerzburg.de}}
}

\vspace*{5mm}
\begin{center}
$^a${\em Physik--Department, Technische Universit\"at M\"unchen}\\
{\em James--Franck--Strasse, D--85748 Garching, Germany}\\
~\\
$^b${\em Department of Theoretical Physics, School of Engineering Sciences}\\
{\em Royal Institute of Technology (KTH) -- AlbaNova University Center}\\
{\em Roslagstullsbacken 21, 106~91 Stockholm, Sweden}\\
~\\
$^c${\em Department of Physics, Oklahoma State University}\\
{\em Stillwater, OK 74078, USA}
\end{center}

\vspace*{20mm}

\centerline{\bf Abstract}
\vspace*{2mm}
We consider discretized gravity in six dimensions, where the two extra
dimensions have been compactified on a hyperbolic disk of constant
curvature. We analyze different realizations of lattice gravity on the disk at the level of an effective field theory for massive
gravitons. It is shown that the observed strong coupling scale of lattice
gravity in discretized five-dimensional flat or warped space can be
increased when the latticized fifth dimension is wrapped around a
hyperbolic disk that has a non-trivial warp factor. As an application, we also study
the generation of naturally small Dirac
neutrino masses via a discrete volume suppression mechanism and
discuss briefly collider implications of our model.

\renewcommand{\thefootnote}{\arabic{footnote}}
\setcounter{footnote}{0}

\newpage

\section{Introduction}\label{sec:Introduction}
Theories with gravity in a curved space-time background exhibit a number of intriguing features. The warped metric
of the Randall-Sundrum (RS) models \cite{Randall:1999ee,Randall:1999vf}, for
example, allows to address the gauge hierarchy problem by considering
our four-dimensional (4D) world as a sub-manifold of a five-dimensional (5D)
anti de Sitter space (alternatively,
see Ref.~\cite{Arkani-Hamed:1998rs}). Apart from that, the AdS/CFT
correspondence \cite{Maldacena:1997re} suggests in warped geometries
an amazing gauge-gravity duality. Moreover, in string
compactifications with fluxes, strongly warped
regions in space-time offer a wide range of new phenomenological and
cosmological applications and serve here, in particular, as a useful device for moduli stabilization \cite{Grana:2005jc}.

A strong space-time curvature is also beneficial
when formulating lattice gravity as an effective field theory
(EFT) \cite{Arkani-Hamed:2002sp,Arkani-Hamed:2003vb,Schwartz:2003vj} (for
related work see, {\it e.g.}, Ref.~\cite{Boulanger:2000rq}). Although lattice gravity
has, to lowest order in its interactions, a strong analogy to the
gauge theory case
\cite{Arkani-Hamed:2001ca,Hill:2000mu} it shows a radically
different strong coupling behavior at the non-linear level. In 5D flat
space, for instance, the ultraviolet (UV) strong
coupling scale depends on the bulk volume, or infrared (IR) scale, via a
so-called ``UV/IR connection'' that would forbid to take the large volume limit
within a sensible EFT \cite{Arkani-Hamed:2003vb}. This UV/IR
connection can, however, be avoided for discretized gravity in 5D
warped space-time \cite{Randall:2005me,Gallicchio:2005mh} (for warped non-gravitational extra dimensions see, {\it e.g.}, Refs.~\cite{Falkowski:2002cm,Bhattacharya:2005xa,Cheng:2001nh}), which has been shown to work in the manifestly holographic regime and admits to
take the large volume limit.

In this paper, we consider a six-dimensional (6D) lattice gravity setup, where the two extra
dimensions are compactified on a discretized hyperbolic disk.
We first analyze a coarse-grained latticization of the hyperbolic disk
and show
that the UV strong coupling scale is essentially set by the
inverse disk radius and becomes independent of the total number of
lattice sites on the boundary. It turns out that, even in the large volume limit, all bounds
from laboratory experiments, astrophysics, and cosmology on
Kaluza-Klein (KK) gravitons are avoided whereas the model would still be testable at a collider. In a second example, we study a fine-grained latticization of the
hyperbolic disk with nonzero warping. We
determine approximately the complete mass spectrum and the
wave-function profiles of the gravitons and estimate the local strong
coupling scale on the disk. We find that the
presence of the 6th dimension can yield the theory on the boundary
sub-manifold of
the disk more weakly coupled than in the corresponding
5D warped case with a single latticized extra dimension. We also analyze the action of a bulk right-handed neutrino in the coarse-grained model
and show that a large curvature of the disk allows to generate
small active Dirac neutrino masses via a volume suppression mechanism.

The paper is organized as follows. In Sec.~\ref{sec:continuum}, we
introduce the extra-dimensional hyperbolic disk that is then
coarsely discretized in Sec.~\ref{sec:coarsegrained}, where we also
discuss collider implications of the model. Next, in Sec.~\ref{sec:effectivetheory}, we calculate the strong coupling
scale for the coarse-grained latticization. In
Sec.~\ref{sec:refined}, we analyze the EFT for a fine-grained
latticization with nonzero warping and demonstrate that
the local strong coupling scale can be pushed to higher values when
going from five to six dimensions. As an application, we present in
Sec.~\ref{sec:neutrinomasses} a mechanism for small Dirac neutrino masses
in the coarse-grained model. Finally, in the Appendix, we derive the
Fierz-Pauli Lagrangian on the hyperbolic disk.

\section{6D warped hyperbolic space}\label{sec:continuum}
Let us consider 6D general relativity compactified to four dimensions
on an orbifold $K_2/Z_2$, where $K_2$ is a two-dimensional hyperbolic disk of constant
negative curvature. We
use capital Roman indices $M=0,1,2,3,5,6$ to label the 6D
coordinates $x^{M}$ while Greek indices $\mu=0,1,2,3$ denote the usual 4D coordinates
$x^{\mu}$. The
6D Minkowski metric is given by
$\eta_{MN}=\textrm{diag}(1,-1,\dots,-1)$. A point on the hyperbolic disk $K_2$ is described by a pair of polar coordinates $(r,\varphi)$, where
$r=x^{5}$ and $\varphi=x^{6}$ are the radial and the
angular coordinates of the point, respectively; on $K_2$, the coordinates $r$
and $\varphi$ can take the values $r\in[0,L]$, where $L\ge 0$
and $\varphi\in[0,2\pi)$. We choose the coordinates such that $r$ is
the geodesic or proper distance of the point $(r,\varphi)$ from the
center, {\it i.e.}, $L$ is the hyperbolic or proper radius of the disk.
The orbifold
$K_2/Z_2$ is obtained from $K_2$ by imposing on the angle $\varphi$
the orbifold projection $\varphi\rightarrow -\varphi$, which restricts
the physical space on the disk to $(r,\varphi)\in [0,L]\times
[0,\pi]$.

\begin{figure}
\begin{center}\includegraphics*[bb = 220 615 390 750, height=5cm]{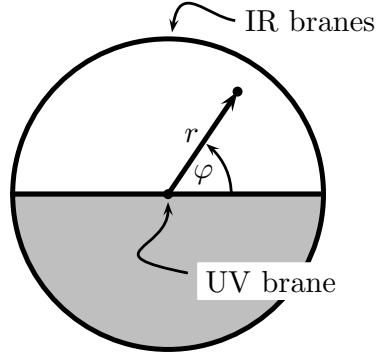}
\end{center}
\vspace*{-5mm}
\caption{{\small Physical space of the orbifold $K_2/Z_2$. Shown are the
coordinates $(r,\varphi)$ of a point on the hyperbolic disk $K_2$. The orbifold projection $\varphi\rightarrow-\varphi$
identifies the upper half of $K_2$ with the lower half (gray
shaded region). The disk is warped along the radial direction
with the UV and IR branes located at the center and at the
boundary of the disk, respectively.}}
\label{fig:orbifold}
\end{figure}

We assume that the disk is warped along the radial
direction like in the 5D RS scenario and that the radial coordinate $r$
takes the role of the 5th dimension in the RS model. The 6D metric
for the warped hyperbolic disk $\tilde{g}_{MN}$ is given by the line element
\begin{equation}
\textrm{d}s^{2}=e^{2\sigma(r)}g_{\mu\nu}(x,r,\varphi)\textrm{d}x^{\mu}\textrm{d}x^{\nu}-\textrm{d}r^{2}-\frac{1}{v^2}\,
\textrm{sinh}^2(v\cdot r)\textrm{d}\varphi^{2},\label{eq:6D-metric}\end{equation}
where $1/v$ is the curvature radius of the disk with $v>0$,
$g_{\mu\nu}(x,r,\varphi)$ is the 4D metric with $x\equiv(x^\mu)$, and
\begin{equation}
\sigma(r)=-w\cdot r,
\end{equation}
where $w$ is the curvature scale for the warping along the radial
direction. We denote the components of the metric $\tilde{g}_{MN}$ in
Eq.~(\ref{eq:6D-metric}) by $\tilde{g}_{\mu\nu}=
e^{2\sigma(r)}g_{\mu\nu}(x,r,\varphi)$,
$\tilde{g}_{55} = -1$, and $\tilde{g}_{66} = -v^{-2}\textrm{sinh}^2(vr)$.
This metric is defined for any values of $r\in [0,L]$, where the
radius $L$ is finite but can be arbitrarily large.
Like in RS I \cite{Randall:1999ee}, we have assumed in
Eq.~(\ref{eq:6D-metric}) orbifold boundary conditions with respect to $r$
at the center $r=0$ and at the boundary of the disk at $r=L$. While we identify
the UV brane with the center at $r=0$, we have IR branes residing at the orbifold fixed points on the boundary at
$r=L$. The orbifold $K_2/Z_2$ with the definition of the coordinates and the
location of the UV and IR branes is shown in Fig.~\ref{fig:orbifold}.
In these coordinates, a concentric circle through a point
$(r,\varphi)$ has a
proper radius $r$ and a proper circumference
$2\pi v^{-1}\textrm{sinh}(vr)$. The proper area of the corresponding disk is
\begin{equation}\label{eq:diskarea}
A=\int \textrm{d}r\,\textrm{d}\varphi\sqrt{|\tilde{g}_{66}|}=4\pi
v^{-2}\sinh^2{(vr/2)}.
\end{equation}
For $vr\gg 1$, the circumference and the area of the circle thus grow
exponentially with $r$.

\begin{figure}
\begin{center}\includegraphics*[height=7cm]{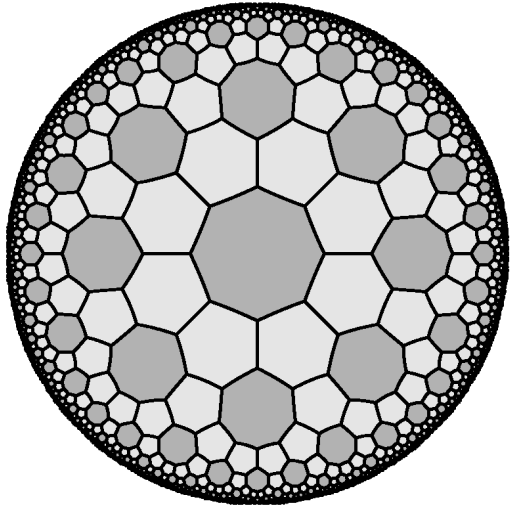}
\end{center}
\vspace*{-5mm}
\caption{{\small Poincar\'e hyperbolic disk with semi-regular $\{6,6,8\}$
tessellation \cite{Sremcevic}.}}
\label{fig:668}
\end{figure}

It is interesting to compare our hyperbolic space with the Poincar\'e
hyperbolic disk. Introducing a new radial coordinate $\hat{r}$ that
is defined by
$\textrm{d}r=(1-\frac{1}{4}v^2\hat{r}^2)^{-1}\textrm{d}\hat{r}$,
we find from Eq.~(\ref{eq:6D-metric}) the Poincar\'e hyperbolic metric
\begin{equation}
\textrm{d}s^{2}=e^{2\sigma(\hat{r})}g_{\mu\nu}(x,\hat{r},\varphi)\textrm{d}x^{\mu}\textrm{d}x^{\nu}-
\frac{1}{(1-\frac{1}{4}v^2\hat{r}^2)^2}(\textrm{d}\hat{r}^2+\hat{r}^2\textrm{d}\varphi^2),\label{eq:Poincaremetric}
\end{equation}
where $\sigma(\hat{r})$ is obtained by replacing in $\sigma(r)=-wr$ the coordinate $r$ by $r(\hat{r})$. The metric in
Eq.~(\ref{eq:Poincaremetric}) has a coordinate singularity at
$\hat{r}=2/v$ corresponding to spatial infinity~$r\rightarrow\infty$. The coordinate $\hat{r}$ is thus restricted to the
interval $\hat{r}\in[0,2/v)$ whereas
$\varphi\in[0,2\pi)$. The polar coordinates $(\hat{r},\varphi)$
define the Poincar\'e hyperbolic disk, which is shown in
Fig.~\ref{fig:668} for the example of a semi-regular $\{6,6,8\}$ tessellation.
In Eq.~(\ref{eq:Poincaremetric}), we have
\begin{equation}
\sigma(\hat{r})=-\frac{w}{v}\,\textrm{log}\left(\frac{1+\frac{v}{2}\hat{r}}{1-\frac{v}{2}\hat{r}}\right).
\end{equation}
In Fig.~\ref{fig:668}, the proper radius and proper circumference of a concentric circle
through a point $(\hat{r},\varphi)$ are $-\sigma(\hat{r})/w$ and
$2\pi\hat{r}/(1-\frac{1}{4}v^2\hat{r}^2)$, which both diverge when
approaching spatial infinity at $\hat{r}=2/v$.
In the rest of the paper, we will, however, always use the coordinates
$(r,\varphi)$ and take the metric as given in Eq.~(\ref{eq:6D-metric}).

On the hyperbolic disk, we write the gravitational action as
\begin{equation}\label{eq:6D-action}
\mathcal{S}=M_{6}^{4}\int\textrm{d}^{6}x\,\sqrt{|\tilde{g}|}(\tilde{R}-2\Lambda),
\end{equation}
where $\tilde{g}\equiv\det\tilde{g}_{MN}$ and it is implied  that
the integration over $r$ and $\varphi$ is restricted to the hyperbolic
disk. In
Eq.~(\ref{eq:6D-action}), $M_{6}$ is the 6D Planck scale, $\Lambda$ the
bulk cosmological constant, and $\tilde{R}=\tilde{g}^{MN}\tilde{R}_{MN}$ is
the 6D curvature scalar, where $\tilde{R}_{MN}$ denotes the 6D Ricci
tensor. The relevant parts of the action can be written in the form
\begin{eqnarray}\label{eq:curved}
\mathcal{S} & = &
M_{6}^{4}\int\textrm{d}^{6}x\sqrt{|g|}\Big[e^{2\sigma(r)}R_{\rm 4D}
-\frac{1}{4}\tilde{g}^{55}\partial_{r}(e^{2\sigma(r)}g_{\mu\nu})(g^{\mu\nu}g^{\alpha\beta}-g^{\mu\alpha}g^{\nu\beta})\partial_{r}(e^{2\sigma(r)}g_{\alpha\beta})
\nonumber\\
&& \qquad\qquad\qquad\quad\:\,
-\frac{1}{4}\tilde{g}^{66}\partial_{\varphi}(e^{2\sigma(r)}g_{\mu\nu})(g^{\mu\nu}g^{\alpha\beta}-g^{\mu\alpha}g^{\nu\beta})\partial_{\varphi}(e^{2\sigma(r)}g_{\alpha\beta})\Big],
\label{eq:splitaction}
\end{eqnarray}
where we have defined $|g|\equiv e^{-8\sigma(r)}|\tilde{g}|$ and
$R_{\rm 4D}$ is the 4D curvature scalar with respect to the
4D metric $g_{\mu\nu}$ (for a definition of $R_{\rm 4D}$ and a derivation
of Eq.~(\ref{eq:curved}), see Appendix \ref{sec:Action-disk}).

We expand $g_{\mu\nu}$ in terms of small fluctuations around 4D
Minkowski space\footnote{We also have $g^{\mu\nu}=\eta^{\mu\nu}-h^{\mu\nu}+h^{\mu\kappa}{h_{\kappa}}^{\nu}+\mathcal{O}(h_{\mu\nu}^{3})$
and~$h^{\mu\nu}\equiv\eta^{\mu\alpha}\eta^{\nu\beta}h_{\alpha\beta}$.}
by replacing $g_{\mu\nu}=\eta_{\mu\nu}+h_{\mu\nu}$. To quadratic
order, the kinetic part of the graviton
Lagrangian density is then, after partial integration, explicitly found to be of the Fierz-Pauli form
\cite{'tHooft:1974bx,Fierz:1939ix}
\begin{eqnarray}\label{eq:linearized}
\mathcal{S}_\textrm{lin}&=& M_6^4\int\textrm{d}^6x\,v^{-1}
\textrm{sinh}(vr)\Big[
\frac{1}{4}e^{2\sigma(r)}
(\partial^{\mu}h^{\nu\kappa}\partial_{\mu}h_{\nu\kappa}-\partial^{\mu}h\partial_{\mu}h-2h^{\mu}h_{\mu}+2h^{\mu}\partial_{\mu}h)\nonumber\\
&&\qquad\qquad\qquad\qquad+\frac{1}{4}e^{4\sigma(r)}
(\partial_{r}h_{\mu\nu})(\eta^{\mu\nu}\eta^{\alpha\beta}-\eta^{\mu\alpha}\eta^{\nu\beta})(\partial_rh_{\alpha\beta})\nonumber\\
&&\qquad\,\,+\frac{1}{4}e^{4\sigma(r)}v^2\textrm{sinh}^{-2}(vr)
(\partial_\varphi
h_{\mu\nu})(\eta^{\mu\nu}\eta^{\alpha\beta}-\eta^{\mu\alpha}\eta^{\nu\beta})(\partial_\varphi
h_{\alpha\beta})\Big],
\end{eqnarray}
where $h={h^{\mu}}_{\mu}$ and $h_{\nu}=\partial^{\mu}h_{\mu\nu}$.
In the analysis of the 5D model in Ref.~\cite{Arkani-Hamed:2003vb},
it has proven useful to neglect the 4D vector, scalar, and radion degrees
of freedom. For simplicity, we have adopted here a similar
gauge where we set $h_{5M}=h_{6M}=0$, for $M=0,1,\dots,6$.

Let us now consider the 6D Einstein tensor
$\tilde{G}_{MN}=\tilde{R}_{MN}-\frac{1}{2}\tilde{g}_{MN}\tilde{R}$ for a flat
4D background $g_{\mu\nu}=\eta_{\mu\nu}$, in which case $\tilde{G}_{MN}$
is on diagonal form
$\tilde{G}_{MN}=\textrm{diag}(\tilde{G}_{11},\tilde{G}_{22},\dots,\tilde{G}_{66})$.
Up to corrections of the order $e^{-vr}$, the components of
$\tilde{G}_{MN}$ asymptote in the IR very quickly to
$\tilde{G}_{11}=\tilde{G}_{22}=\tilde{G}_{33}=-\tilde{G}_{00}=(3vw-6w^2-v^2)e^{-2wr}$, $\tilde{G}_{55}=4vw-6w^2$,
and $\tilde{G}_{66}=-10w^2v^{-2}\textrm{sinh}^2(vr)$. In this limit, the 6D Einstein equations
can be formally satisfied by (i) adding to the gravitational action in
Eq.~(\ref{eq:6D-action}) a ``matter''
action
\begin{equation}\label{eq:6D-action-n_AB}
\mathcal{S}_\textrm{m}=M_{6}^4\int
\textrm{d}^6x\sqrt{|\tilde{g}|}\tilde{g}^{AB}n_{AB},
\end{equation}
where $n_{AB}$ denotes the tensor
$n_{AB}\equiv -\tilde{G}_{AB}$, and by (ii) setting the bulk
cosmological constant $\Lambda$ in Eq.~(\ref{eq:6D-action}) equal to
$\Lambda=-\frac{1}{2}\tilde{G}^A_{\,\,\,A}$. Note that for zero warping $w=0$, we
obtain the exact result that the tensor $n_{AB}$ becomes constant
throughout the bulk.

In what follows, we will view discrete gravitational extra dimensions as a tool to
explore general relativity without the need to explicitly solve Einstein's
equations \cite{Randall:2005me}. Here, the massive gravitons in the
compactified theory can be understood in terms of a graph consisting
of sites and links called ``theory space'', and it is not necessary to
find, {\it e.g.}, the gravitational source for an action of the type in
Eq.~(\ref{eq:6D-action-n_AB}).

\section{Coarse-grained discretization}\label{sec:coarsegrained}
In this section, we consider a discretization of the hyperbolic disk
$K_2$ with a metric as given in Eq.~(\ref{eq:6D-metric}). In particular, we analyze
here the case where the warping along the radial
direction is set to zero, {\it i.e.}, it is $w=0$, while the curvature
radius $v$ is non-vanishing. The general case, with both $w\neq 0$ and $v\neq 0$, will be studied later in Sec.~\ref{sec:refined}.
\subsection{Latticization}\label{sec:coarsediscretization}
\begin{figure}
\begin{center}\includegraphics*[bb = 125 585 285 735]{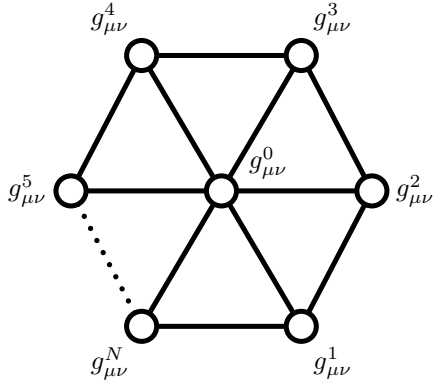}\end{center}
\vspace*{-5mm}
\caption{{\small Coarse-grained
discretization of the hyperbolic disk $K_2$. The circles represent the
sites and are labeled as $i=0,1,2,\dots ,N$. On each site $i$ lives a
graviton field $g_{\mu\nu}^i$ and two neighboring sites $i$
and $j$ are connected by a link $(i,j)_\textrm{link}$ (solid lines).}}
\label{fig:disk}
\end{figure}

Let us consider the coarse-grained latticization of the hyperbolic
disk $K_2$ that is described by the diagram in Fig.~\ref{fig:disk}.
The diagram is spanned by $N$ sites located on the
boundary of $K_2$, which are labeled as $i=1,2,\dots,N$ with the identification $i+N\equiv i$, while the center of the disk is
represented by a single site $i=0$. The sites on the boundary of $K_2$ are evenly spaced
on a concentric circle with proper radius $L$ around the central site
$i=0$ residing at the origin, {\it i.e.}, the $i$th site on the
boundary has polar coordinates $(r,\varphi)=(L,i\cdot\Delta\varphi)$, where
$\Delta\varphi=2\pi/N$.
In Fig.~\ref{fig:disk}, two sites $i$ and $i+1$ on the boundary
are connected by a link $(i,i+1)_\textrm{link}$ while the site $i=0$ in the
center is connected to all $N$ sites on the boundary by the links
$(0,i)_\textrm{link}$. To implement a lattice gravity theory in terms of this triangulation,
we will throughout interpret the sites and links following the theory of massive gravitons in
Refs.~\cite{Arkani-Hamed:2002sp,Arkani-Hamed:2003vb,Schwartz:2003vj}.
Here, each site $i$ is equipped with its own metric
$g^{i}_{\mu\nu}$, which can be expanded around flat space as
$g^i_{\mu\nu}=\eta_{\mu\nu}+h^i_{\mu\nu}$, where $\eta_{\mu\nu}$ is
the usual 4D Minkowski metric.
In a naive latticization of the linearized action $\mathcal{S}_\textrm{lin}$ in
Eq.~(\ref{eq:linearized}), we then replace the derivatives on the sites as
\begin{equation}\label{eq:derivatives}
\partial_\varphi h_{\mu\nu}\rightarrow\frac{1}{\Delta\varphi}(
h_{\mu\nu}^{i+1}-h_{\mu\nu}^i),\quad
\partial_r h_{\mu\nu}\rightarrow\frac{1}{L}(h^i_{\mu\nu}-h_{\mu\nu}^0),
\end{equation}
which gives the Fierz-Pauli graviton mass terms\footnote{For a recent discussion of Fierz-Pauli mass terms and ghosts in massive
gravity see Ref.~\cite{Creminelli:2005qk}.}
\begin{eqnarray}
\mathcal{S}_{\textrm{FP}} &  = &
M_{4}^{2}\int\textrm{d}^{4}x\,\sum_{i=1}^{N}
\left[m_{*}^{2}(h_{\mu\nu}^{i}-h_{\mu\nu}^{0})(\eta^{\mu\nu}\eta^{\alpha\beta}-\eta^{\mu\alpha}\eta^{\nu\beta})(h_{\alpha\beta}^{i}-h_{\alpha\beta}^{0})\right.\nonumber \\
&
&\qquad\qquad\qquad\left.+m^{2}(h_{\mu\nu}^{i+1}-h_{\mu\nu}^{i})(\eta^{\mu\nu}\eta^{\alpha\beta}-\eta^{\mu\alpha}\eta^{\nu\beta})(h_{\alpha\beta}^{i+1}-h_{\alpha\beta}^{i})\right],\label{eq:FP-masses}\end{eqnarray}
where $M_4$ is the ``local'' universal 4D Planck scale on each of the
sites whereas $m_*$ and $m$ respectively denote the proper inverse
lattice spacings in radial and angular direction. To relate $M_4$ to
$M_6$, note that Eq.~(\ref{eq:splitaction}) yields upon discretization
\begin{equation}\label{eq:discretization}
M_6^4\int \textrm{d}^6 x\,\sqrt{|\tilde{g}|}\rightarrow \frac{M_6^4 A}{N} \sum_{i=0,1}^N \int \textrm{d}^4 x\,\sqrt{|g_4|},
\end{equation}
where~$A$ is the proper area of the disk from Eq.~(\ref{eq:diskarea})
with~$r=L$ and~$g_4=\textrm{det}\,g_{\mu\nu}$. In
Eq.~(\ref{eq:discretization}), it is understood that the sum starts
from~$i=0$ for the kinetic terms and from~$i=1$ for the mass terms,
respectively. The mass scales in Eq.~(\ref{eq:FP-masses}) can then be
matched as
\begin{equation}\label{eq:matching}
m_{*}=\frac{1}{L},\quad m=\frac{Nv}{2\pi\,\textrm{sinh}(vL)},
\quad M_4^2=\frac{M_6^{4}A}{N},\quad
M_{\rm Pl}^2=M_6^4A=M_4^2N,
\end{equation}
where $M_{\rm Pl}\simeq 10^{18}\:{\rm GeV}$ is the usual 4D Planck
scale of the low-energy theory. In the basis
$(h_{\mu\nu}^{0},h_{\mu\nu}^{1},\dots,h_{\mu\nu}^{N})$,
the $(N+1)\times(N+1)$ graviton mass matrix reads
\begin{equation}
M_{\textrm{g}}^{2}=m_{*}^{2}\left(\begin{matrix}N & -1 & -1 & \cdots & -1\\
-1 & 1\\
-1 &  & 1\\
\vdots &  &  & \ddots\\
-1 &  &  &  & 1\end{matrix}\right)+m^{2}\left(\begin{matrix}0 & 0 & 0 & \dots & 0\\
0 & 2 & -1 &  & -1\\
0 & -1 & 2 & \ddots\\
\vdots &  & \ddots & \ddots & -1\\
0 & -1 &  & -1 & 2\end{matrix}\right),\label{eq:Mg}\end{equation}
where the blank entries are zero. Diagonalization
of $M_{\textrm{g}}^{2}$ leads to the graviton mass
spectrum \begin{equation}
M_{0}^{2}=0,\quad M_{n}^{2}=m_{*}^{2}+4m^{2}{\textrm{sin}}^{2}\frac{\pi n}{N},\quad M_{N}^{2}=(N+1)m_{\ast}^{2},\label{eq:eigenvalues}\end{equation}
where $n=1,2,\dots,N-1$. Note that the spectrum in Eq.~(\ref{eq:eigenvalues})
has been described previously for the gauge theory case \cite{Bauer:2003mh,Hallgren:2004mw}.
Denoting the $n$th mass eigenstate with mass $M_{n}$ as given in Eq.~(\ref{eq:eigenvalues}) by $\hat{H}_{\mu\nu}^{n}$,
the mass eigenstates can be written as \begin{subequations}\label{eq:eigenstates}
\begin{eqnarray}
\hat{H}^0_{\mu\nu}&=&\frac{1}{\sqrt{N+1}}(1,1,1,\dots ,1),\label{eq:zeromode}\\
\hat{H}^n_{\mu\nu}&=&\frac{1}{\sqrt{N}}(0,1,e^{{\textrm{i}}\frac{2n\pi}{N}},
e^{{\textrm{i}}\frac{4n\pi}{N}},\dots,e^{{\textrm{i}}\frac{2(N-1)n\pi}{N}}),
\label{eq:nthmode}\\
\hat{H}^N_{\mu\nu}&=&\frac{1}{\sqrt{N(N+1)}}(N,-1,-1,\dots, -1),\label{eq:Nthmode}
\end{eqnarray}
\end{subequations} where in Eq.~(\ref{eq:nthmode}) the index $n$ runs over $n=1,2,\dots,N-1$
and we have, for definiteness, taken $N$ to be even. The mass eigenstates
include a zero mode $\hat{H}_{\mu\nu}^{0}$ with flat profile. Note that
the $N-1$ massive eigenstates $\hat{H}_{\mu\nu}^{n}$ in Eq.~(\ref{eq:nthmode})
are all exactly located on the boundary while the $N$th massive mode
$\hat{H}_{\mu\nu}^{N}$ in Eq.~(\ref{eq:Nthmode}) is peaked
in the center. The tower of masses $M_{n}$ belonging to the states
$\hat{H}_{\mu\nu}^{n}$ reproduces for $n\ll N$
a linear KK-type spectrum $\sim m\cdot n/N$ that is
sitting on top of $m_{\ast}^{2}$. From Eq.~(\ref{eq:matching}), we
also observe that the limit $m\ll m_{\ast}$ can be realized for~$N\gg
1$ with a sufficiently large value of~$vL$. In this case, the $N-1$ states in Eq.~(\ref{eq:nthmode}) become practically
degenerate with mass $m_{*}$ and the $N$th
mode $\hat{H}_{\mu\nu}^{N}$ becomes very heavy, {\it i.e.}, $M_N\gg m_{\ast}$.

\subsection{Phenomenological implications}
It has been recently demonstrated that large extra dimensions can
be hidden in ``multi-throat'' configurations where the occurrence of KK
gravitons is delayed to energies that are much higher than the inverse
radius of the extra dimension \cite{Kim:2005aa}. Such a multi-throat
geometry is found in our coarse-grained model as a special
case in the limit $m\rightarrow0$. Therefore, by the same arguments as in
Ref.~\cite{Kim:2005aa}, the corrections to Newton's law in the coarse-grained model become only important when the
distance between two test masses is shorter than a critical radius
$r_{\textrm{crit}}\simeq m_{*}^{-1}\:{\textrm{log}}\: N$. For $N
\simeq 10^{30}$, {\it e.g.}, we could thus have, like in large extra dimensions, the local Planck scale on the sites lowered down to
$M_4\simeq M_\textrm{Pl}/\sqrt{N}\simeq 1\:\textrm{TeV}$ while
for $m_{*}\gtrsim100\:{\textrm{MeV}}$ all current bounds on large
extra dimensions from Cavendish-type experiments \cite{Hoyle:2004cw}, astrophysics
\cite{Hannestad:2001xi}, and cosmology \cite{Hall:1999mk} are avoided.

The collider implications for the direct production of KK gravitons
$\hat{H}^n$ in $e^{+}+e^{-}\rightarrow \gamma + \hat{H}^{n}$ and
$p+p\rightarrow {\rm jet}+\hat{H}^{n}$ reactions can be determined
like in the large extra dimensional scenario:
Standard Model (SM) matter located on a single site $k$ on the boundary of the
disk interacts with gravity as
\begin{equation}\label{eq:interaction}
\mathcal{S}_{\rm int} \approx \frac{1}{M_\text{Pl}}\int d^{4}x\,
T^{\mu\nu}\Big(\sum_{n=1}^{N-1}e^{-{\rm i}2\pi k\cdot
n/N}\hat{H}^{n}_{\mu\nu}+
\hat{H}^{0}_{\mu\nu}+\frac{1}{\sqrt{N}}\hat{H}^{N}_{\mu\nu}\Big)+{\rm
h.c.},
\end{equation}
where $T^{\mu\nu}$ is the SM stress-energy tensor. We have in this
expression for large $N$ used the approximation $N+1\approx N$. We see
that SM matter has $1/M_\textrm{Pl}$ suppressed couplings to the
graviton zero mode and to the $N-1$ quasi-degenerate modes, whereas
the couplings to the heavy mode are even suppressed by a factor
$1/\sqrt{N}M_{\rm Pl}$. For example, assuming $N=10^{30}$ sites and
inverse lattice spacings $m_{*}=14\:\textrm{TeV}$ and $m =
0.14\:\textrm{GeV}$, we find for a center-of-momentum (CoM) energy
$E_\textrm{CoM}=14.1\:\textrm{TeV}$ and a quark energy $E_{\rm q}\gtrsim
10^2\:\textrm{GeV}$ that the cross section for the production of an
individual KK graviton via the parton subprocess $q+g\rightarrow
q+\hat{H}^n$ is $\sigma \sim 10^{-30}\:\textrm{pb}$. This is, of
course, not observable at a collider (as a detectability requirement
of signal above background we take $\sigma \gtrsim 1-100\:\textrm{fb}$
\cite{Giudice:1998ck}). But summing over the high multiplicity of
$N=10^{30}$ states, we actually have here a total cross section
$\sigma_{\rm total}\sim 1\:\textrm{pb}$, which could give a signal at
the LHC. The signal would be a jet plus missing transverse energy. Of
course, a proper treatment of the reaction $p+p\rightarrow {\rm
jet}+\hat{H}^{n}$ should take into account the summation over the
parton subprocesses with the appropriate parton distribution
functions, which is, however, beyond the scope of this paper.

\begin{figure}
\begin{center}\includegraphics*[height=7cm]{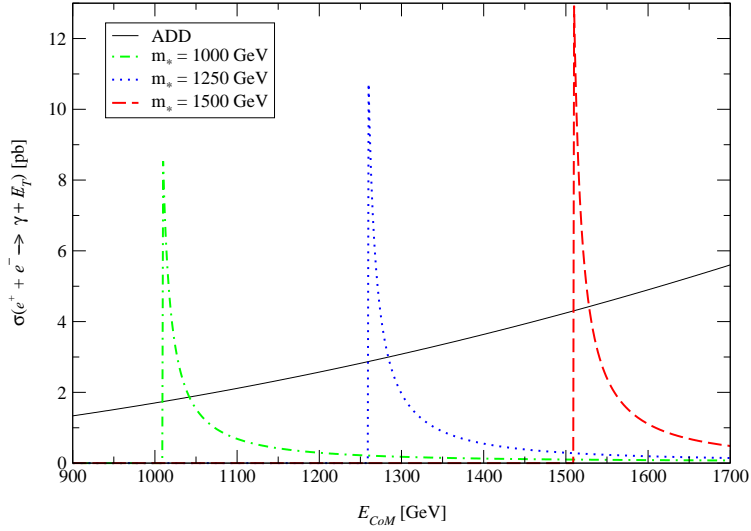}
\end{center}
\vspace*{-5mm}
\caption{{\small Cross section $\sigma(e^{+}+e^{-}\rightarrow \gamma +
\Slash{E}_{T})$ in units of pb versus $E_{{\rm CoM}}$ in GeV. Shown is
the cross section for the coarse-grained model with $m_{*}=1000$
GeV (dash-dotted), $m_{*}=1250$ GeV (dotted), and $m_{*}=1500$ GeV
(dashed). In all cases, we have chosen $m=1$ GeV and $N=10^{30}$
sites. For comparison, we have also presented a typical ADD model with 2 extra
dimensions and $M_{f}=2.5$ TeV (solid line). We have also made a kinematic cut by
taking $E_{\gamma}\gtrsim 10$ GeV. The coarse-grained model gives a signal that is
qualitatively different from usual large extra dimensions. The sharp peak
close to $E_{{\rm CoM}}\sim m_{*}$ is due to the nearly degenerate
spectrum located around $E_{{\rm CoM}}\sim m_{*}$. Below $E_{{\rm
CoM}}\sim m_{*}$, the cross section drops to zero because of the
mass-gap in the disk model.}}
\label{fig:crosssection}
\end{figure}

For a choice of parameters $N=10^{30}$, $m_{*}=1.0\:\textrm{TeV}$, $m
=1.0\:\textrm{GeV}$, $E_{\rm CoM}=1.01\:\textrm{TeV}$, and a lower
cutoff for the photon energy $E_{\gamma}\gtrsim 10\:\textrm{GeV}$,
the cross section for the production of an individual KK mode via
$e^{+}+e^{-}\rightarrow \gamma + \hat{H}^n$ is $\sigma \sim
10^{-29}\:\textrm{pb}$, which results in a total cross section of the
order $\sigma_{\rm total}\sim 8\:\textrm{pb}$ and could, {\it
e.g.}, be observable at the ILC. The signal would be a photon plus
missing transverse energy.  In Fig.~\ref{fig:crosssection}, we present
the cross sections for this reaction as a function of $E_{{\rm
CoM}}$. We give the results for $m_{*}=1000$ GeV, $m_{*}=1250$
GeV, and $m_{*}=1500$ GeV. In all cases, we have chosen $N=10^{30}$,
$m=1$ GeV, and a lower cutoff for the photon energy $E_{\gamma}\gtrsim 10\:\textrm{GeV}$. We have also made a comparison
with the ADD scenario \cite{Arkani-Hamed:1998rs} for 2 large extra dimensions and a fundamental scale
$M_{f}=2.5$ TeV. The coarse-grained model has a qualitatively different
behavior compared to large extra dimensions. The sharp peak close to
$E_{\rm CoM}\sim m_{*}$ is due to the nearly degenerate spectrum
located around $E_{\rm CoM}\sim m_{*}$. Below $E_{\rm CoM}\sim m_{*}$,
the cross section is zero because of the large mass gap between the
zero mode and the first excited mode.

In addition, the exchange of virtual
KK modes could influence the cross sections for some SM processes
which we will, however, not discuss in further detail here.

\section{Effective theory}\label{sec:effectivetheory}
In this section, we will study the strong coupling behavior of the
coarse-grained model presented in Sec.~\ref{sec:coarsediscretization}, where we use
throughout the EFT for massive gravitons introduced in
Refs.~\cite{Arkani-Hamed:2002sp,Arkani-Hamed:2003vb,Schwartz:2003vj}.
Before turning to the complete latticized disk, however, it is instructive
to consider first the strong coupling scales in two of its
sub-geometries.
The first one is represented by the boundary of the disk, which we
will call the ``circle'' geometry. The second one is obtained by removing
the links on the boundary, and we will henceforth call this theory
space the ``star'' geometry (for studies of related geometries, see,
{\it e.g.}, Ref.~\cite{star}). First, in Secs.~\ref{sec:circle},
\ref{sec:star}, and \ref{sec:disk}, we determine the strong coupling
scale by restricting to the Goldstone boson sector. Then, in
Sec.~\ref{sec:coarsematter}, we discuss the strong coupling scale that is
actually seen by an observer localized on a single site on the boundary of the disk.

\subsection{Strong coupling in the circle geometry}\label{sec:circle}
To implement the EFT on the latticized hyperbolic disk $K_2$, we
replace in Eq.~(\ref{eq:derivatives}) the differences
$h^i_{\mu\nu}-h^j_{\mu\nu}$ between the graviton fields on two sites $i$
and $j$ that are connected by a link $(i,j)_\textrm{link}$ according to
\begin{subequations}\label{eq:restoringGCs}
\begin{equation}\label{eq:differences}
h_{\mu\nu}^i-h_{\mu\nu}^j\,\,\rightarrow\,\,
g_{\mu\nu}^i-\partial_{\mu}Y_{i,j}^\alpha\partial_{\nu}Y_{i,j}^\beta
g^{j}_{\alpha\beta},
\end{equation}
where $Y_{i,j}^{\mu}$ denotes a link field that can be written as
\begin{equation}
Y_{i,j}^{\mu}(x_{\mu})=x^{\mu}+A_{i,j}^{\mu}(x_{\mu})+\partial^{\mu}\phi_{i,j}(x_{\mu}),\label{eq:links}\end{equation}
\end{subequations}
in which $A_{i,j}^{\mu}$ and $\phi_{i,j}$
represent the vector and scalar components of the Goldstone bosons that
restore general coordinate invariances in the EFT. For a detailed
description of the technique for restoring general coordinate invariances, see Refs.~\cite{Arkani-Hamed:2002sp,Arkani-Hamed:2003vb,Schwartz:2003vj}.

The circle geometry is obtained from the coarse-grained latticized
disk in Sec.~\ref{sec:coarsediscretization} by assuming
$m_*=0$ and $m$ nonzero. Since this reduces the model to the 5D lattice
gravity theory discussed in Ref.~\cite{Arkani-Hamed:2003vb}, we will
only give a short review of this case here.

The kinetic Lagrangian of the gravitons is
found by specializing in Eqs.~(\ref{eq:restoringGCs}) the links $(i,j)_\textrm{link}$
to the case $(i,j)_\textrm{link}=(i,i-1)_\textrm{link}$, where $i=2,\ldots ,N,N+1$. As a
consequence, we obtain from the Fierz-Pauli mass terms in
Eq.~(\ref{eq:FP-masses}) for the circle geometry the action
\begin{equation}\label{eq:circle}
\mathcal{S}_{\rm circle}=\mathcal{S}_{\rm FP}+
\int {\rm d}^4x\:M_4^2
\sum_{i=1}^N(h^i_{\mu\nu}\Box
h^i_{\mu\nu}
+m^2[(h^{i}_{\mu\nu}-h^{i-1}_{\mu\nu})\Box
\phi_{i,i-1}+(\Box\phi_{i,i-1})^3]).
\end{equation}
The kinetic mixing between the gravitons and scalar Goldstones can be eliminated
by partial integration and a subsequent Weyl rescaling, which
produces proper kinetic terms for the Goldstones
\cite{Arkani-Hamed:2002sp}. As shown in
Ref.~\cite{Arkani-Hamed:2003vb}, this EFT for massive gravitons has a cutoff
\begin{equation}\label{eq:5Dstrongcoupling}
\Lambda_{\rm circle}=(N^{3/2}M_4m_1^4)^{1/5}=(M_{\rm Pl}m^4/N^3)^{1/5},
\end{equation}
where~$m_1=m/N$ is the mass of the lightest massive graviton, and we have used that the usual 4D Planck scale $M_{\rm
Pl}\simeq 10^{18}\:{\rm GeV}$ is related to the local 4D Planck scale on
the sites $M_4$ by $M_{\rm Pl}=\sqrt{N}M_4$ [see Eq.~(\ref{eq:matching})]. The strong coupling scale
$\Lambda_{\rm circle}$ depends on
the number of lattice sites $N$, {\it i.e.}, on the circumference
$R=Nm^{-1}$ of the circle. The fact that the UV scale $\Lambda_{\rm
circle}$ depends on the IR scale $R$ has been called UV/IR
connection and restricts in a sensible EFT the possible number
of sites to a maximal value $N_{\rm max}=(M_4^2mR^3)^{1/8}$
\cite{Arkani-Hamed:2003vb}.

\subsection{Strong coupling in the star geometry}\label{sec:star}
The star geometry arises from the coarse-grained disk model
by setting the parameters $m=0$ and $m_*\neq 0$. Note that this geometry is
similar to the multi-throat configuration which has been
analyzed for the gauge theory case in
Ref.~\cite{Kim:2005aa}. In contrast to Sec.~\ref{sec:circle}, we specialize here to the
case in which the links $(i,j)_\textrm{link}$ in
Eqs.~(\ref{eq:restoringGCs}) are
$(i,j)_\textrm{link}=(i,0)_\textrm{link}$, where $i=1,2,\dots, N$. We
then get from the Fierz-Pauli mass terms
in Eq.~(\ref{eq:FP-masses}) the action
\begin{equation}\label{eq:star}
\mathcal{S}_{\rm star}=\mathcal{S}_{\rm FP}+
\int {\rm d}^4x\:M_4^2
(h^0_{\mu\nu}\Box h^0_{\mu\nu}+\sum_{i=1}^N(h^i_{\mu\nu}\Box
h^i_{\mu\nu}
+m_*^2[(h^{0}_{\mu\nu}-h^i_{\mu\nu})\Box\phi_{0,i}+(\Box\phi_{0,i})^3])),
\end{equation}
where we have also included the kinetic terms of the gravitons.
In Eq.~(\ref{eq:star}), the terms
$m_*^2(h_{\mu\nu}^0-h_{\mu\nu}^i)\Box\phi_{0,i}$ can be written as
a $(N+1)\times N$ kinetic matrix which is proportional to
\begin{equation}\label{eq:kineticmatrix1}
\left(
\begin{matrix}
1 & 1 & 1 & \dots & 1\\
-1 & 0 & 0 & \dots & 0\\
0 &-1 & 0 & \dots & 0\\
0 & 0 &-1 & \dots & 0\\
\vdots & \vdots & \vdots & \ddots & \vdots\\
0 & 0 & 0 & \dots & -1
\end{matrix}
\right),
\end{equation}
where the rows and columns are spanned by
$(h_{\mu\nu}^0,h_{\mu\nu}^1,\dots,h_{\mu\nu}^N)$ and
$(\phi_{0,1},\phi_{0,2},\dots,\phi_{0,N})$, respectively.
To determine here the strong coupling
scale, we express the gravitons on the boundary as the linear combinations
$h^i_{\mu\nu}=\frac{1}{\sqrt{N}}\sum_{n=1}^Ne^{{\rm i}2\pi i\cdot
n/N}{H'}^{n}_{\mu\nu}$, where we have introduced the fields
${H'}^n_{\mu\nu}=H_{\mu\nu}^n$, for $n=2,3,\dots ,N-1$, with
$H_{\mu\nu}^n$ from Eq.~(\ref{eq:nthmode}) while
${H'}^N_{\mu\nu}$ is in the basis of Eqs.~(\ref{eq:eigenstates}) given
by the $(N+1)$-component vector
${H'}^N_{\mu\nu}=\frac{1}{\sqrt{N}}(0,1,1,\dots ,1)$.
Similarly, we expand the scalar components of
the Goldstone fields belonging to the links $(i,0)_\textrm{link}$ as
$\phi_{i,0}=\frac{1}{\sqrt{N}}\sum_{n=1}^Ne^{{\rm i}2\pi i\cdot
n/N}\Phi_n$. The matrix in Eq.~(\ref{eq:kineticmatrix1}) gets then
transformed to
\begin{equation}\label{eq:kineticmatrix2}
\left(
\begin{matrix}
0 & 0 & 0 & \dots & \sqrt{N}\\
-1 & 0 & 0 & \dots & 0\\
0 &-1 & 0 & \dots & 0\\
0 & 0 &-1 & \dots & 0\\
\vdots & \vdots & \vdots & \ddots & \vdots\\
0 & 0 & 0 & \dots & -1
\end{matrix}
\right),
\end{equation}
where the rows and columns have been written in the bases
$(h^0_{\mu\nu},{H'}^1_{\mu\nu},{H'}^2_{\mu\nu},\dots
,{H'}^N_{\mu\nu})$ and $(\Phi_1,\Phi_2,\dots,\Phi_{N})$,
respectively. Let us next perform a rotation by the angle
${\rm arctan}(\sqrt{N})$ acting in the 2-dimensional subspace
$(h_{\mu\nu}^0,{H'}^N_{\mu\nu})$ from
the left on the first and last rows of the matrix in
Eq.~(\ref{eq:kineticmatrix2}). This rotates away the entry $\sqrt{N}$
in the top-right corner and introduces new basis vectors via
$(h^0_{\mu\nu},{H'}^N_{\mu\nu})\rightarrow(H^0_{\mu\nu},H^N_{\mu\nu})$,
with $H^0_{\mu\nu}$ and $H^N_{\mu\nu}$ as defined in Eqs.~(\ref{eq:zeromode})
and (\ref{eq:nthmode}), respectively. We can then approximate the action in
Eq.~(\ref{eq:star}) by
\begin{eqnarray}\label{eq:momentumbasis}
\mathcal{S}_{\rm star}&=&\mathcal{S}_{\rm FP}+
\int {\rm d}^4x\:M_4^2
\Big[h^0_{\mu\nu}\Box h^0_{\mu\nu}+\sum_{n=1}^N\Big(H^n_{\mu\nu}\Box
H^{-n}_{\mu\nu}\nonumber\\
&-&m_*^2H_{\mu\nu}^n\Box\Phi_{-n}\Big)
+\sum_{n,m=1}^N\Big(\frac{m_*^2}{\sqrt{N}}(\Box\Phi_n)(\Box\Phi_m)(\Box\Phi_{-n-m})\Big)\Big],
\end{eqnarray}
where we have defined $H_{\mu\nu}^{-n}=H_{\mu\nu}^{N-n}$ and $\Phi_{-n}=\Phi_{N-n}$.
From this equation, we read off the graviton fields in
canonical normalization
$\hat{H}^n_{\mu\nu}=M_4H_{\mu\nu}^n$, and, consequently, the canonically normalized scalar Goldstone modes are
$\hat{\Phi}_n=M_4m_*^2\Phi_n$. In Eq.~(\ref{eq:momentumbasis}), we
thus have
\begin{equation}
M_4^2\frac{m_*^2}{\sqrt{N}}(\Box\Phi_n)(\Box\Phi_m)(\Box\Phi_{-n-m})
=\frac{1}{\sqrt{N}M_4m_*^4}
(\Box\hat{\Phi}_n)(\Box\hat{\Phi}_m)(\Box\hat{\Phi}_{-n-m}).
\end{equation}
From the amplitude for $\hat{\Phi}_n-\hat{\Phi}_m$ scattering, we hence find
for the star geometry the strong coupling scale
\begin{equation}\label{eq:strongcoupling}
\Lambda_{\rm star}=(\sqrt{N}M_4m_*^4)^{1/5}=(M_{\rm Pl}m_*^4)^{1/5},
\end{equation}
which is independent from the number of sites $N$ on the boundary. This is different from the circle geometry since the discrete radial derivatives in Eq.~(\ref{eq:star}) do not depend on~$N$.
Therefore, the UV/IR connection problem from the circle geometry is absent in
the star geometry. Note that the strong coupling
scale in Eq.~(\ref{eq:strongcoupling}) appears to be what we would
naively expect in the theory of a single massive graviton with mass $m_*$
\cite{Arkani-Hamed:2002sp}. However, as we will see later in
Sec.~\ref{sec:coarsematter}, the strong coupling
scale actually seen by a 4D brane-localized observer on the boundary
is in the limit $m_\ast\rightarrow M_4$ smaller than in the theory of a single massive graviton.

\subsection{Strong coupling on the disk}\label{sec:disk}
In the disk model of Sec.~\ref{sec:coarsediscretization}, the links form a mesh which would generally imply the presence
of additional uneaten pseudo-Nambu-Goldstone bosons.
These bosons, however, can acquire large masses from invariant plaquette
terms that are added to the action \cite{Arkani-Hamed:2002sp}.
In our model, we will therefore assume that these scalars
decouple and estimate the strong coupling scale on the
disk $\Lambda_{\rm disk}$ only from the scattering of the modes
that are eaten by the gravitons.

\begin{figure}
\begin{center}
\includegraphics*[bb = 80 640 480 720]{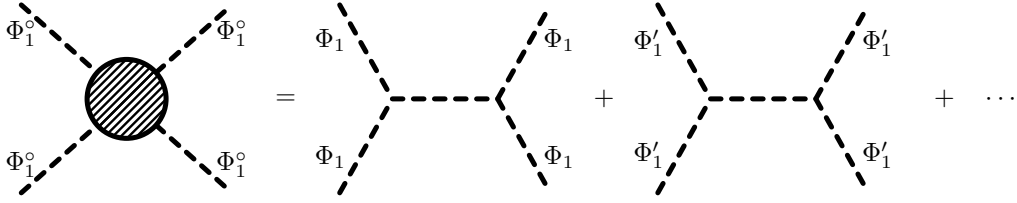}
\end{center}
\vspace*{-5mm}
\caption{\small Contributions to the scattering amplitude of the
lightest massive graviton in the coarse-grained disk model. For large $N$, the scalar
$\Phi^\circ_1$ is dominantly composed of $\Phi_1$, and the relative contribution from $\Phi'_1-\Phi'_1$ scattering
becomes suppressed by a factor $\sim 1/N^6$.}\label{fig:scattering}
\end{figure}

The combination of $\mathcal{S}_{\rm circle}$ and
$\mathcal{S}_{\rm star}$ gives, using the notation of
Sec.~\ref{sec:star}, in momentum basis the total action of the disk
\begin{eqnarray}\label{eq:Sdisk}
\mathcal{S}_{\rm disk}&=&\mathcal{S}_{\rm FP}+
\int {\rm d}^4x\:M_4^2
\Big[ h^0_{\mu\nu}\Box h^0_{\mu\nu}+\sum_{n=1}^N\Big(H^n_{\mu\nu}\Box
H^{-n}_{\mu\nu}\nonumber\\
&-&H_{\mu\nu}^n\Box(m_*^2\Phi_{-n}+m^2(1-e^{-\textrm{i}2\pi\cdot
n/N})\Phi'_{-n})\Big)
\nonumber\\
&+&\sum_{n,k=1}^N\Big(\frac{m_*^2}{\sqrt{N}}(\Box\Phi_n)(\Box\Phi_k)(\Box\Phi_{-n-k})
+\frac{m^2}{\sqrt{N}}(\Box\Phi'_n)(\Box\Phi'_k)(\Box\Phi'_{-n-k})\Big)\Big],
\end{eqnarray}
where $\Phi_n$ and $H^n_{\mu\nu}$ are defined as in
Eq.~(\ref{eq:momentumbasis}) while the $\Phi'_n$ are related to the
scalar Goldstones $\phi_{i,i-1}$ of the circle geometry by
$\phi_{i,i-1}=\frac{1}{\sqrt{N}}\sum_{n=1}^Ne^{{\rm i}2\pi i\cdot
n/N}\Phi'_n$. From Eq.~(\ref{eq:Sdisk}), we hence find that the Goldstones $\Phi_n^\circ$ which become the scalar longitudinal
components of the gravitons $H_{\mu\nu}^n$ are, for $n\ll N$, approximately given by
\begin{equation}\label{eq:wouldbeGoldstone}
\Phi_n^\circ\approx m_\ast^2\Phi_n+2\pi\frac{n}{N}m^2\Phi'_{n}.
\end{equation}
Thus, for large $N$, the field $\Phi_n^\circ$ becomes dominantly composed of
$\Phi_n$, whereas $\Phi'_n$ makes up almost entirely the uneaten linear
combination orthogonal to $\Phi^\circ_n$. Consider now in
Eq.~(\ref{eq:Sdisk}) the tri-linear derivative coupling term involving
the field $\Phi'_n$. In the pure circle geometry discussed in
Sec.~\ref{sec:circle}, the coupling of this term becomes strong for large
$N$ and is responsible for the UV/IR connection. In our disk model,
however, the linear combination orthogonal to $\Phi^\circ_n$
decouples. In this limit, since the admixture of $\Phi'_n$ to
$\Phi_n^\circ$ is only of the order $\sim 1/N$, the tri-linear
coupling of $\Phi'_n$ in Eq.~(\ref{eq:Sdisk})
picks up a suppression factor $\sim 1/N^3$, when expressed in terms
of the canonically normalized fields $\hat{\Phi}^\circ_n=M_4\Phi_n^\circ$, and takes approximately the form
\begin{equation}
\frac{m^2}{\sqrt{N}}(\Box\Phi'_n)(\Box\Phi'_k)(\Box\Phi'_{-n-k})
\rightarrow
\frac{m^8}{N^{7/2}M_4m_\ast^{12}}(\Box\hat{\Phi}^\circ_n)(\Box\hat{\Phi}^\circ_k)(\Box\hat{\Phi}^\circ_{-n-k}),
\end{equation}
where we have assumed that $n,k\sim 1 \ll N$. Unless $m$ is not too large
compared to $m_\ast$, the dominant contribution to the massive
graviton scattering is therefore given by the tri-linear derivative
coupling of the fields $\Phi_n$ in Eq.~(\ref{eq:Sdisk}) (see Fig.~\ref{fig:scattering}).
This coupling is
similar to the term in Eq.~(\ref{eq:momentumbasis}) that sets the
strong coupling scale in the star geometry. For $N\rightarrow\infty$, the strong coupling scale of the disk model
$\Lambda_{\rm disk}$ thus converges to
\begin{equation}\label{eq:diskstrongcoupling}
\Lambda_{\rm disk}\rightarrow\Lambda_{\rm star}=(M_{\rm Pl}m_\ast^4)^{1/5}.
\end{equation}
The EFT for massive gravitons in our disk model does therefore not
suffer from the UV/IR connection of the circle sub-geometry and thus
allows to take the large $N$ limit. Again, the strong
coupling scale in Eq.~(\ref{eq:diskstrongcoupling}) has been estimated
by considering the self-interactions of the Goldstone bosons
only. Next, in Sec.~\ref{sec:coarsematter}, we will take the coupling
to matter into account and determine the strong coupling scale that is
actually observed on a single site on the boundary of the disk.

\subsection{Inclusion of matter}\label{sec:coarsematter}
\begin{figure}
\begin{center}
\includegraphics*[bb = 160 630 470 720]{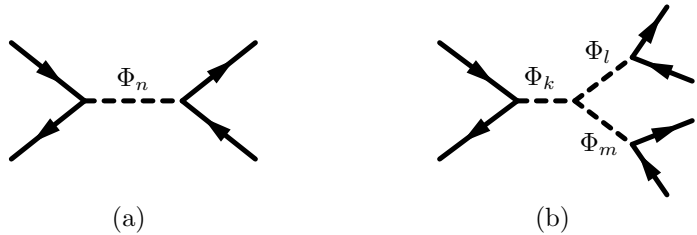}
\end{center}
\vspace*{-5mm}
\caption{\small Scattering of SM fermions via Goldstone modes. In the
diagrams (a) and (b), we have $n,k=0,1,2,\dots,N$,
and $k=l+m\;\text{mod}\,N$. The high multiplicity of internal states
[$N$ for (a) and $N^2$ for (b)] introduces an upper bound on the actually observed strong coupling scale of the
order of the local Planck scale $\lesssim M_4$.}\label{fig:matterscattering}
\end{figure}

The strong coupling scales given for the star and the disk model in Eqs.~(\ref{eq:strongcoupling}) and
(\ref{eq:diskstrongcoupling}) are, of course, deceptive, since they
have been derived by looking at the Goldstone boson sector alone. To find
the actual strong coupling scale as seen by an observer, we have
to take into account the coupling of the Goldstone modes to SM matter. In
what follows, we shall therefore consider SM matter located at a
single site $k$ of the boundary, which interacts with gravity via the action $\mathcal{S}_\text{int}$ as
given in Eq.~(\ref{eq:interaction}). After an appropriate Weyl
rescaling (for details see
Refs.~\cite{Arkani-Hamed:2002sp,Gallicchio:2005mh}),
$\mathcal{S}_\text{int}$ leads to the following interaction between the
Goldstone bosons and matter:
\begin{equation}
\mathcal{S}^\Phi_\text{int}\approx \frac{1}{M_\text{Pl}}\int
d^4x\Big(\sum_{n=1}^{N-1}e^{-{\rm i}2\pi k\cdot n/N}\hat{\Phi}_{-n} T
+\frac{1}{\sqrt{N}}\hat{\Phi}_0 T\Big)+ {\rm h.c.},
\end{equation}
where $T=\text{Tr}(T_{\mu\nu})$ and we have approximated $N+1\approx
N$. Note that the coupling of $T$ to an individual Goldstone boson
$\hat{\Phi}_n$ is suppressed by a factor $1/M_\text{Pl}$, but the sum
over the high multiplicity of Goldstones can lower the strong coupling
scales estimated in Secs.~\ref{sec:star} and \ref{sec:disk}
significantly. SM fermions, for example, see strong coupling effects
via processes of the types shown in Fig.~\ref{fig:matterscattering}.
From dimensional analysis, we estimate for the diagrams (a) and (b) in
Fig.~\ref{fig:matterscattering} that the strong coupling scales as seen by
a SM observer become roughly
\begin{equation}\label{eq:strongcouplings}
\text{Diagram (a)}\::\:\Lambda_\text{disk}\rightarrow M_4,\qquad
\text{Diagram (b)}\::\:\Lambda_\text{disk}\rightarrow \sqrt{M_4 m_\ast}.
\end{equation}
The actual strong coupling scale is therefore lower than that of the
theory of a single massive graviton. However, since the strong
coupling scales in Eq.~(\ref{eq:strongcouplings}) are, for a fixed
local Planck scale $M_4$, independent from the number of sites $N$ on the boundary, the UV/IR
connection problem is still absent -- even after taking into account
the high multiplicity of Goldstone bosons that couple to SM
matter. When taking the limit $m_\ast\rightarrow M_4$, we find for
diagram (b) in Eq.~(\ref{eq:strongcouplings}) that the strong
coupling scale seen by an observer becomes
$\Lambda_\text{disk}\rightarrow M_4$, that means, the actually
observed strong coupling scale $\Lambda_\text{disk}$ can be as large
as the local Planck scale $M_4$ on the sites of the boundary.

\section{Fine-grained discretization}\label{sec:refined}
In this section, we will consider a refinement of the
discretization of the hyperbolic disk described in
Sec.~\ref{sec:coarsegrained}. Different from our earlier discussion,
however, we will now assume that the disk is strongly warped along the
radial direction and consider finally also a nonzero warping in angular direction.

\subsection{Structure of the graph}\label{sec:structure}
\begin{figure}
\begin{center}
\includegraphics*[bb = 200 520 400 700, height=7.0cm]{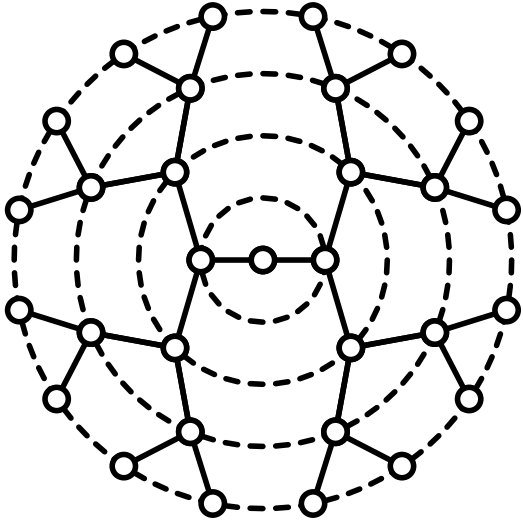}
\end{center}
\vspace*{-5mm}
\caption{\small Fine-grained discretization of the hyperbolic disk. The
graph shows the special case $\ell=2$ and $k_{\rm max}=4$. Angular
links are represented as dashed lines and radial links as solid lines.}\label{fig:tessellation}
\end{figure}

In choosing a refined discretization of the hyperbolic disk,
we will, in what follows, confine ourselves to the class of graphs that are collectively
represented by Fig.~\ref{fig:tessellation}, where we interpret the sites and
links as in Sec.~\ref{sec:coarsediscretization}.
The sites are placed on
concentric circles (dashed lines) around a site in the center of the disk.
Starting from the innermost circle and going outward, the circles are labeled
as $k=1,2,\dots ,k_{\rm max}$, where the outermost circle,
labeled by $k_{\rm max}$, is identified with the boundary of the disk. Fig.~\ref{fig:tessellation} depicts the special case $k_{\rm max}=4$.

Let us now specify the structure of the graph in more detail. We distinguish between two types of links:
``angular'' and ``radial'' links. In Fig.~\ref{fig:tessellation}, the
angular links connect neighboring sites on the same circle and are
shown as dashed curved lines; the radial links connect neighboring
sites that are not on the same circle and are drawn as solid straight lines.

The site in the center is connected with $\ell$ radial links to the
sites on the first circle. Every site on the $k$th circle, where
$1\leq k<k_{\rm max}$, is connected by $\ell$ radial links to the nearest
$\ell$ sites on the next outer circle with label $k+1$. In
Fig.~\ref{fig:tessellation}, we have the special case $\ell=2$.
To maximize the symmetry of the graph,
we assume that all sites on a given circle are evenly spaced and that the links emanating outward from any given site on some
inner circle with label $k<k_{\rm max}$ are always symmetrically arranged
with respect to a 2-fold mirror symmetry axis through this site and the
center of the disk. In what follows, we will, for simplicity and unless otherwise
mentioned, specialize to the case $\ell =2$ as in
Fig.~\ref{fig:tessellation} but leave $k_{\rm max}$ a free positive integer
parameter. The generalization to $\ell > 2$ is then straightforward.
On the $k$th circle, we hence have $N_k\equiv 2^k$ sites, and the
total number of sites on the first $k$
circles is $N_k^{\rm total}\equiv 2^{k+1}-1$.  For later notational
convenience, we denote by $N_0^{\rm total}=N_0=1$ the number of sites
(only one) in the center. We wish to emphasize that the exponential growth
of lattice sites, when moving from the center outward on the disk, is a
salient feature of our graph that will play an important role in the
study of the EFT in Sec.~\ref{sec:local}.

Similar to the 5D case in Ref.~\cite{Randall:2005me}, we assume that the radial geodesic coordinates of the sites
are integer multiples of a common proper radial lattice spacing $a_r$ and take
the values
\begin{equation}\label{eq:ar}
r_k=k\cdot a_r,
\end{equation}
where $k=0,1,2,\dots, k_\textrm{max}$, {\it i.e.}, the $k$th
circle has a proper radius $r_k$. We define $m_*\equiv 1/a_r$ as the
inverse radial lattice spacing between two adjacent circles.
Furthermore, since the sites on each given circle are evenly spaced, we define the
lattice spacing in $\varphi$ direction between two neighboring sites on the $k$th
circle as $a_{\varphi,k}=2\pi\,\textrm{sinh}(vr_k)/(vN_k)$,
where $2\pi\,\textrm{sinh}(vr_k)/v$ is
the proper circumference of the $k$th circle. The inverse lattice
spacing in $\varphi$ direction on the
$k$th circle will then be
written as $m_k\equiv 1/a_{\varphi,k}$, which is in general $k$-dependent.

To simplify further the discussion as much as possible, we will, in what follows, go to an
approximation in which all radial links have the same universal proper
length $a_r$. This is, of course, an arbitrarily good approximation
for the set of
radial links that radiate outward from the $k$th circle when $k\gg 1$
and $vL\ll 1$. However, in our approximation, we even assume
that all the $N_k^\textrm{total}-1$ radial links on the disk have a
common proper length $a_r$, irrespective of their position on the disk
or the size of the curvature radius.

\subsection{Radial warping}\label{sec:subgraph}
Let us now consider the implementation of a nonzero warping along the
radial direction of the refined discretized hyperbolic disk introduced in
Sec.~\ref{sec:structure}. The graph we are concerned with is of the type
given in Fig.~\ref{fig:tessellation}, where the number of
circles $k_{\rm max}$ can be arbitrary.
In this section, we will, for simplicity,
neglect the effect of the angular links (dashed lines) and assume that
the graviton masses are generated only by the radial links (solid
lines). As we will show in Sec.~\ref{sec:angular}, the angular links
can be easily taken into account in full generality for the complete
disk model and do not have significant impact on the relevant properties of
the model with radial links only.

Each site on the disk can be specified by an index pair
$(k,j)$, where $k$ is the label of the circle where the site is
located and $j=1,2,\dots,N_k$ labels ({\it e.g.}, in clockwise
direction) all the sites on this circle. The site in the center
carries the index
pair $(0,1)$. The metric living on the site $(k,j)$ is
written as $g^{k,j}_{\mu\nu}=\eta_{\mu\nu}+h^{k,j}_{\mu\nu}$,
where $h^{k,j}_{\mu\nu}$ is the graviton field on this site. We adopt
a similar notation in momentum space and denote the canonically
normalized graviton mass
eigenstates by $\hat{H}^{k,j}_{\mu\nu}$, where $k=1,2,\dots,k_{\rm max}$ and
$j=1,2,\dots ,N_k$ while $\hat{H}_{\mu\nu}^{0,1}$ is the zero mode. The
mass eigenvalue belonging to $\hat{H}^{k,i}_{\mu\nu}$ will be denoted by $\lambda_{k,i}$.

\begin{figure}
\begin{center}
\includegraphics*[bb = 190 520 450 700, height=6.0cm]{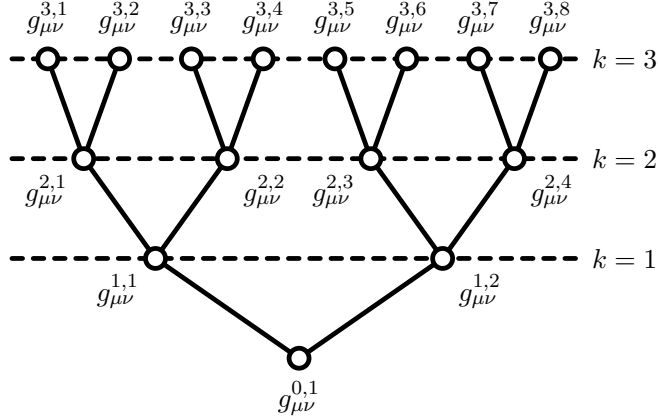}
\end{center}
\vspace*{-5mm}
\caption{\small Sub-graph of the discretized hyperbolic
disk obtained by truncating the disk at $k_\textrm{max}=3$.
The horizontal dashed lines represent segments from the concentric circles
with labels $k=1,2,3$. The labeling of the sites as $(k,i)$ is made
explicit for the associated metrics $g_{\mu\nu}^{k,i}$ (see
text).}\label{fig:tree}
\end{figure}

To specify the labeling of the sites more exactly,
consider Fig.~\ref{fig:tree}, which is another way of
representing the inner part of the disk in
Fig.~\ref{fig:tessellation}, where the center appears now at the bottom of the graph and the
concentric circles are shown as dashed horizontal lines.
The radial links are again drawn as straight solid lines.
In the above notation, any given site $(k,i)$ on the
$k$th circle, where $k=0,1,2,\dots,k_\textrm{max}-1$, is connected via two
radial links with two nearest neighboring sites $(k+1,j_1)$ and
$(k+1,j_2)$ on the next outer circle. The indices $j_1$ and $j_2$ are
recursively defined as functions of the index $i$ by setting
$j_1=2(i-1)+1$ and $j_2=2(i-1)+2$. Starting with the label $(0,1)$ for the center,
this prescription fixes completely the assignment of labels
for all sites on the circle by requiring the
indices $i$ in $(k,i)$ to be ordered, {\it e.g.}, in
clockwise direction on the disk. The labeling is shown explicitly in
Fig.~\ref{fig:tree} for the fields
$g_{\mu\nu}^{k,i}$ living on the sites $(k,i)$.

To introduce the warping along the radial direction of the fine-grained latticized disk, we proceed exactly like in the discretized 5D RS model
in Ref.~\cite{Randall:2005me} and replace in the
linearized gravitational action in Eq.~(\ref{eq:linearized}), {\it e.g.}, the
derivatives in the $r$ direction by
\begin{equation}\label{eq:warping}
e^{-2wr}\partial_r
h_{\mu\nu}\rightarrow\frac{1}{a_r}e^{-2wka_r}
(h_{\mu\nu}^{k,j}-h_{\mu\nu}^{k-1,i}),
\end{equation}
where it is understood that the sites $(k-1,i)$ and $(k,j)$, with $k=1,2,\dots ,k_\textrm{max}$,
are nearest neighbors on adjacent circles connected by a radial
link. It is convenient to introduce a dimensionless number
$\epsilon<1$, which is related to the warp factor by $\epsilon\equiv e^{-wa_r}$.
Since from each site in the interior of the disk exactly two radial
links are pointing outward, the index $j$ in Eq.~(\ref{eq:warping}) can be either $j=2(i-1)+1$ or
$j=2(i-1)+2$. We thus obtain for the discretized action on the warped hyperbolic disk
\begin{equation}\label{eq:Swarp}
\mathcal{S}_\textrm{lin}=
\int {\rm d}^4x\:M_4^2
\sum_{k=0}^{k_\textrm{max}}\sum_{i=1}^{N_k}\epsilon^{2k}h^{k,i}_{\mu\nu}\Box
h^{k,i}_{\mu\nu}+\mathcal{S}_\textrm{FP},
\end{equation}
where $\mathcal{S}_\textrm{FP}$ are the Fierz-Pauli mass terms
\begin{eqnarray}\label{eq:warpedFP}
\mathcal{S}_{\textrm{FP}} &  = &
M_{4}^{2}\int\textrm{d}^{4}x\,\sum_{k=1}^{k_{\rm max}}
\sum_{i=1}^{N_{k-1}}
m_{*}^{2}\epsilon^{4k}\Big[(h_{\mu\nu}^{k-1,i}-h_{\mu\nu}^{k,n_i})(\eta^{\mu\nu}\eta^{\alpha\beta}-\eta^{\mu\alpha}\eta^{\nu\beta})(h_{\alpha\beta}^{k-1,i}-h_{\alpha\beta}^{k,n_i})\nonumber \\
&&\qquad\qquad\qquad+(h_{\mu\nu}^{k-1,i}-h_{\mu\nu}^{k,n_i+1})(\eta^{\mu\nu}\eta^{\alpha\beta}-\eta^{\mu\alpha}\eta^{\nu\beta})(h_{\alpha\beta}^{k-1,i}-h_{\alpha\beta}^{k,n_i+1})\Big],\end{eqnarray}
in which, according to our labeling, the index $n_i$ is given by $n_i=2(i-1)+1$, where
$i=1,2,\dots ,N_{k-1}$. From the kinetic terms in
Eq.~(\ref{eq:Swarp}), we find that $M_k\equiv M_4\epsilon^k$ is the
local Planck scale on each site on the $k$th circle. In the basis of canonically normalized fields
$\hat{h}^{k,i}_{\mu\nu}=M_4\epsilon^{k}h^{k,i}_{\mu\nu}$, we find from
the Fierz-Pauli mass terms that the mass
matrix element between two fields $\hat{h}_{\mu\nu}^{k,i}$ and
$\hat{h}_{\mu\nu}^{l,j}$ is given by
\begin{equation}\label{eq:massmatrix}
M_{(k,i)(l,j)}=\frac{1}{a_r^2}\epsilon^{2(l+1)}[
(2\epsilon^2+\epsilon^{-2})\delta_{k,l}\delta_{i,j}-\epsilon^{-1}\delta_{k+1,l}
(\delta_{n_i,j}+\delta_{n_i+1,j})-\epsilon\,\delta_{k-1,l}\delta_{\lceil
i/2\rceil,j}],
\end{equation}
where it is assumed that $k>0$, while $n_i$ is as in Eq.~(\ref{eq:warpedFP}),
and $\lceil\dots\rceil$ is the ceiling function. The matrix
element $M_{(0,1)(l,j)}$ is obtained by dropping in
Eq.~(\ref{eq:massmatrix}) the quantity $\epsilon^{-2}$. In what follows,
we will work in the ``rough, local flat space approximation'' of
Ref.~\cite{Randall:2005me}, where the product $wa_r$ is moderately small, {\it
i.e.}, $wa_r<1$. In this approximation, the mass matrix in
Eq.~(\ref{eq:massmatrix}) reduces to
\begin{equation}\label{eq:approximation}
M_{(k,i)(l,j)}\approx\frac{1}{a_r^2}\epsilon^{2(l+1)}[(2+\epsilon^{-2})\delta_{k,l}\delta_{i,j}-
\delta_{k+1,l}(\delta_{n_i,j}+\delta_{n_i+1,j})-\delta_{k-1,l}\delta_{\lceil
i/2\rceil,j}],
\end{equation}
where $k>0$. To arrive at Eq.~(\ref{eq:approximation}), we have set in
Eq.~(\ref{eq:massmatrix}) $\epsilon,\epsilon^2=1$ but
kept $\epsilon^{-2}$ and the higher powers $\epsilon^{2l}$ which can be small for $l\gg
1$. The matrix element $M_{(0,1),(l,j)}$ is in this approximation obtained from Eq.~(\ref{eq:approximation}) by dropping $\epsilon^{-2}$.
Note that this mass matrix is
what one would expect in the corresponding gauge theory case
\cite{Falkowski:2002cm,Bhattacharya:2005xa,Cheng:2001nh}.

Let us now comment on the range of parameters we are interested
in. Like in the RS model, we will assume that the usual 4D low energy
Planck scale $M_\textrm{Pl}\simeq 10^{18}\:\textrm{GeV}$ is given by the local Planck scale $M_4$ on
the UV brane, {\it i.e.}, we set $M_4\approx M_\textrm{Pl}$, which holds
approximately as long as $\epsilon\cdot\ell\ll 1$, say, if $\epsilon\cdot\ell\lesssim 0.5$.
For definiteness, we can, {\it e.g.}, take the inverse radial lattice
spacing $m_\ast\simeq (0.05)\times M_4$, assume that the
small expansion parameter is $\epsilon\simeq 0.1$, and suppose that
the curvature scale for the warping $w$ is somewhat smaller than the
fundamental scale, {\it i.e.}, $w\simeq
(0.1)\times M_\textrm{Pl}$. To realize for these parameters, {\it e.g.}, the RS I model with the SM on one of the sites on the boundary of the disk, we need on the boundary a local Planck scale $M_{k_\textrm{max}}=\epsilon^{k_\textrm{max}}M_\textrm{Pl}\simeq
1\:\textrm{TeV}$, which implies that in this case the disk has a
proper radius that is roughly $k_\textrm{max}=\mathcal{O}(10)$ units $a_r$
wide. For our special case with $\ell=2$, we would then have
$N_{k_\textrm{max}}=\mathcal{O}(10^4)$ sites on the boundary.

If we want that $m_\ast$ becomes as large as $m_\ast=M_4\approx M_\textrm{Pl}$,
while keeping the same number of sites on the boundary and
$w\simeq(0.1)\times M_\textrm{Pl}$, we can
``smooth out'' our graph by inserting between each pair of neighboring
circles roughly $20$ extra circles on which the number of sites $N_k$
grows sufficiently slowly when moving outward on the disk. This is achieved by formally replacing in
the above example $\epsilon\rightarrow\epsilon^{1/20}=(0.1)^{1/20}$ and
$\ell\rightarrow\ell^{1/20}=2^{1/20}$, in which case we would have
$N_k=\ell^{k/20}$ and a
proper radius that is
$k_\textrm{max}\simeq\mathcal{O}(10^2)$ units $M_4^{-1}$
wide.

\subsection{Spectrum for a subgraph}\label{sec:subgraphspectrum}
Before analyzing the masses and eigenstates of the gravitons in the
model for an arbitrary number of circles $k_\textrm{max}$, let us
first study the spectrum for the subgraph in Fig.~\ref{fig:tree}. From Eq.~(\ref{eq:approximation}), we find in the basis
$(\hat{h}_{\mu\nu}^{0,1},\hat{h}_{\mu\nu}^{1,1},\hat{h}_{\mu\nu}^{1,2},\hat{h}_{\mu\nu}^{2,1},\hat{h}_{\mu\nu}^{2,2},\hat{h}_{\mu\nu}^{2,3},\hat{h}_{\mu\nu}^{2,4})$
for the graviton mass matrix
\begin{equation}\label{eq:subgraph}
M_g^2=m_\ast^2\epsilon^2
\left(
\begin{matrix}
2 & -1 & -1  & 0 & 0 &
0 & 0\\ -1  & 1 +2\epsilon^2 & 0 &
-\epsilon^2 & -\epsilon^2 & 0 & 0 \\ -1 & 0 &
1 +2\epsilon^2 & 0 & 0 &-\epsilon^2 &
-\epsilon^2\\ 0 & -\epsilon^2 & 0 &
\epsilon^2 & 0 & 0 & 0\\ 0 & -\epsilon^2 & 0
& 0 & \epsilon^2 & 0 & 0\\ 0 & 0 &
-\epsilon^2 & 0 & 0 & \epsilon^2 & 0\\ 0 & 0
& -\epsilon^2 & 0 & 0 & 0 & \epsilon^2\\
\end{matrix}
\right),
\end{equation}
where we have truncated the graph in Fig.~\ref{fig:tree} at the dashed
line labeled by $k=2$. In what follows, we will determine the masses
and mass eigenstates of the gravitons to leading order in
perturbation theory following the example of the warped gauge theory
case in Refs.~\cite{Falkowski:2002cm,Kim:2005aa}.

In the following, we choose a convention where the mass eigenvalue
$\lambda_{k,i}$ belonging to $\hat{H}^{k,i}_{\mu\nu}$ measures the actual mass in multiples of $m_\ast^2$.
The mass eigenstates of the matrix in Eq.~(\ref{eq:subgraph}) contain
a flat zero mode
\begin{subequations}\label{eq:subgraphstates}
\begin{eqnarray}
\hat{H}^{0,1}_{\mu\nu}=\frac{1}{\sqrt{7}}(1,1,1,1,1,1,1)&:&\lambda_{0,1}=0
\label{eq:subgraphzeromode}
\end{eqnarray}
and two heavy modes
\begin{eqnarray}
\hat{H}^{1,1}_{\mu\nu}=\frac{1}{\sqrt{2}}(0,-1,1,0,0,0,0)&:&\lambda_{1,1}=\epsilon^2,\\
\hat{H}^{1,2}_{\mu\nu}=\frac{1}{\sqrt{6}}(-2,1,1,0,0,0,0)&:&\lambda_{1,2}=3\epsilon^2,
\end{eqnarray}
where we have listed in each line also the eigenvalue corresponding to
each eigenstate. Observe that the zero mode has an
exactly flat wave-function profile whereas the two heavy modes are
more localized toward the center of the disk. Furthermore, the
spectrum contains the following three degenerate states
\begin{eqnarray}
\hat{H}^{2,j}_{\mu\nu}=\frac{1}{2}
(0,0,0,1,e^{{\rm i}2\pi j/4},e^{{\rm i}4\pi j/4},e^{{\rm i}6\pi
j/4})&:& \lambda_{2,j}=\epsilon^4,
\end{eqnarray}
where $j=1,2,3$, as well as the slightly heavier mode
\begin{eqnarray}
\hat{H}^{2,4}_{\mu\nu}=\frac{1}{\sqrt{84}}(-4,-4,-4,3,3,3,3)&:&\lambda_{2,4}=\frac{7}{3}\epsilon^4.\label{eq:heavier}
\end{eqnarray}
\end{subequations}
It is important to note that the mode profile in Eq.~(\ref{eq:heavier}) and the flat profile of the zero mode~$\hat{H}^{0,1}_{\mu\nu}$ in Eq.~(\ref{eq:subgraphzeromode}) are only a result
of going in Eq.~(\ref{eq:approximation}) to the rough, local flat space approximation. We will discuss the differences to the exact profiles in the following section.

\subsection{Spectrum for the general case}\label{sec:generalcase}
\begin{figure}
\begin{centering}
\includegraphics[clip,width=1.0\columnwidth]{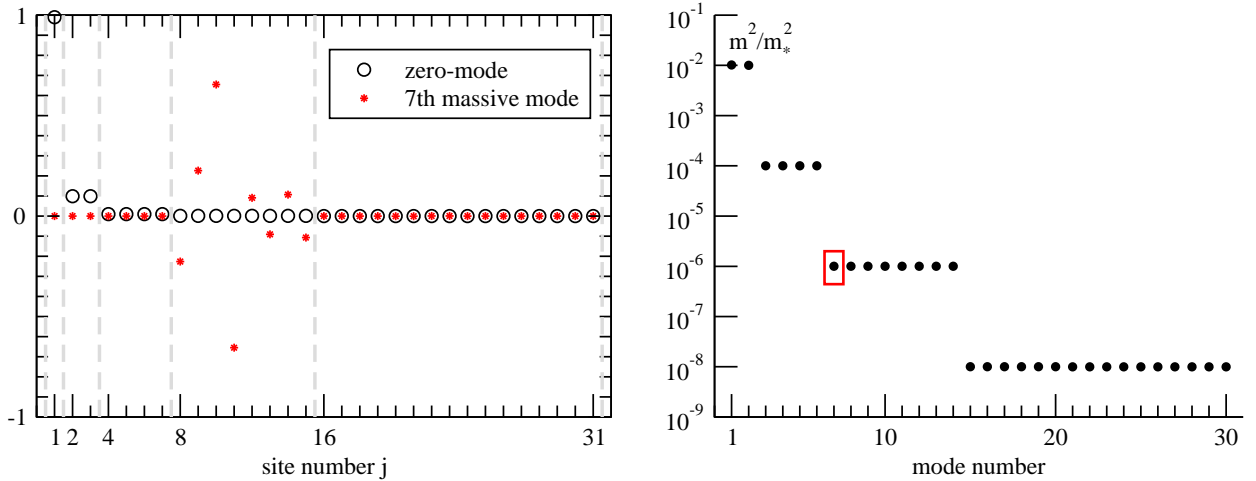}\par
\end{centering}
\caption{\label{fig:Mode07profiles}\small According to the ``exact'' mass
matrix in Eq.~(\ref{eq:massmatrix}), the left plot shows for~$k_{\rm
max}=4$ and~$\epsilon=0.1$ the exact mode profiles of the zero mode
(drawn as circles) and the 7th massive mode (drawn as stars). The
latter would in the rough, local flat space approximation
correspond to the mode in Eq.~(\ref{eq:heavymode}). The
vertical dashed lines separate the sites on each circle, which
illustrates that the zero mode is localized near the UV
brane~($j=1$) and that the 7th mode has support only on the third
circle~($j=8,\dots,15$). The right plot shows the ``exact'' mass
spectrum, where the 7th massive mode is marked by a box and the
zero mode has been omitted.}
\end{figure}

Let us now consider the graviton spectrum for an arbitrary number of
concentric circles $k_\textrm{max}$ in the rough, local flat space approximation of Sec.~\ref{sec:subgraph}. We determine the
eigenvalues and eigenstates of the graviton mass matrix to leading
order in perturbation theory like in
Sec.~\ref{sec:subgraphspectrum}. In this general case, we thus find that the graviton mass
matrix has a flat zero mode
\begin{eqnarray}\label{eq:0mode}
\hat{H}^{0,1}_{\mu\nu}=\frac{1}{\sqrt{N_{k_{\rm max}}^{\rm total}}}(1,1,\ldots,1)&:&\lambda_{0,1}=0
\end{eqnarray}
and two heavy modes
\begin{eqnarray}
\hat{H}^{1,1}_{\mu\nu}= \frac{1}{\sqrt{2}}(0,1,e^{{\rm
i}\pi},0,\ldots,0)&:&\lambda_{1,1}=\epsilon^2,\label{eq:degheavy}\\
\hat{H}^{1,2}_{\mu\nu}=
\frac{1}{\sqrt{6}}(-2,1,1,0,\ldots,0)&:&\lambda_{1,2}=3\epsilon^2.
\label{eq:maxmode}
\end{eqnarray}
In addition, there are for each circle with label $k=2,3,\ldots, k_{\rm max}$ in total
$N_k-1$ degenerate modes that read
\begin{eqnarray}\label{eq:deglight}
\hat{H}^{k,j}_{\mu\nu}=\frac{1}{\sqrt{N_k}}
(\underbrace{0,\ldots,0}_{N_{k-1}^{\rm total}},\underbrace{1,
e^{{\rm i}2\pi j/N_{k}},e^{{\rm i}4\pi j/N_{k}},\ldots, e^{{\rm i}2\pi
j(N_{k}-1)/N_{k}}}_{N_k},0,\ldots,0)&:& \lambda_{k,j}=\epsilon^{2k},
\end{eqnarray}
where $j=1,2,\ldots,N_{k}-1$. These modes are all
located on the $k$th circle. For each $k=2,3,\dots, k_{\rm
max}$, this set of $N_k-1$ states is accompanied by a single slightly
heavier mode which has an eigenvalue
$\lambda_{k,N_k}\approx 2\epsilon^{2k}$ and is approximately given by
\begin{eqnarray}\label{eq:heavymode}
\hat{H}^{k,N_k}_{\mu\nu}=
\frac{1}{\sqrt{N_k^{\rm
  total}}}(\underbrace{-1,\ldots,-1}_{N_{k-1}^{\rm
  total}},\underbrace{1,\ldots,1}_{N_k},0,\ldots,0)
  &:&\lambda_{k,N_k}\approx 2\epsilon^{2k}.
\end{eqnarray}
Note that, to leading order, these modes vanish outside the $k$th
circle. As in Sec.~\ref{sec:subgraphspectrum}, these ``delocalized''
profiles and the flat zero mode $\hat{H}_{\mu\nu}^{0,1}$ in
Eq.~(\ref{eq:0mode}) are only an artifact from working in the rough,
local flat space approximation. In Fig.~\ref{fig:Mode07profiles}, a
numerical calculation shows that the original mass matrix in
Eq.~(\ref{eq:massmatrix}) in fact reproduces a zero mode that is, as
expected from the RS model, localized on the UV brane. Also the
slightly heavier modes in Eq.~(\ref{eq:heavymode}) become actually
localized on single circles. The spectrum and wave-function profiles
of the gravitons for the general case are summarized in Table \ref{tab:spectrum}.
\begin{table}
\begin{center}
\begin{tabular}{|c|c|c|}
\hline${\rm mass}/m_*^2$ & $\#\,{\rm
of}\,{\rm modes}$ & ${\rm wavefunction}$\\\hline\hline
$0$ & $1$ & Eq.~(\ref{eq:0mode}) \\
$\epsilon^2$ & $1$ &
Eq.~(\ref{eq:degheavy})\\
$3\epsilon^2$ & $1$ & Eq.~(\ref{eq:maxmode})\\
$\epsilon^{2k}$ & $N_k-1$ &
Eq.~(\ref{eq:deglight})\\ $\approx 2\epsilon^{2k}$ & $1$ &
Eq.~(\ref{eq:heavymode}) \\\hline
\end{tabular}
\caption{\small Mass spectrum and wave-function profiles for the
gravitons in the rough, local flat space approximation, taking into
account radial links only. Here, $N_k=2^k$ and
$k=2,3,\ldots, k_{\rm max}$.}\label{tab:spectrum}
\end{center}
\end{table}

\subsection{Effect of angular links}\label{sec:angular}
We can include the effect of the angular links on the total graviton
mass matrix in Sec.~\ref{sec:generalcase} by adding to each sub-block spanned by the fields on the $k$th
circle a matrix of the form
\begin{equation}
M^k_{ij}=m_k^2\epsilon^{2k}(2\delta_{i,j}-\delta_{i-1,j}-\delta_{i+1,j}),
\end{equation}
where $i,j=1,2,\dots,N_k$ and $m_k$ is the proper inverse lattice spacing on the $k$th
circle as defined in Sec.~\ref{sec:structure}. The inclusion of
angular links in the fine-grained model is therefore
similar in effect to what happens in the example of the coarse-grained model in
Sec.~\ref{sec:coarsediscretization}. Switching on the angular links
leaves the zero mode unaffected but changes the masses of the
degenerate modes $\hat{H}^{k,j}_{\mu\nu}$ in Eq.~(\ref{eq:deglight}) as
\begin{equation}
m_\ast^2\lambda_{k,j}\rightarrow
m_\ast^2\epsilon^{2k}+4m_k^2\epsilon^{2k}\sin^{2}\frac{j\pi}{N},
\end{equation}
where $j=1,2,\ldots, N_k$. This is the analog of Eq.~(\ref{eq:eigenvalues}).

\subsection{Local strong coupling scale}\label{sec:local}
Let us now consider the local strong coupling scale on the discretized
hyperbolic disk. In determining the strong coupling scale, we will
restrict here ourselves first to the Goldstone boson sector. Later, in
Sec.~\ref{sec:finegrainedmatter}, we will take the coupling of the Goldstone
bosons to matter into account and discuss the strong coupling scale
that is actually experienced by a SM observer localized on a single site on
the boundary of the disk. By the same arguments
as in Sec.~\ref{sec:disk}, we will
restrict ourselves here again to the case where only the effect of the
radial links is taken into account and
the interactions associated with the angular links can be neglected
in the large $N_k$ limit. To analyze the effective theory,
we will use a new notation for the link fields that
is different from the labeling employed previously in Secs.~\ref{sec:coarsediscretization} and
\ref{sec:effectivetheory}. From now on, in the notation of
Sec.~\ref{sec:generalcase}, a radial link field connecting two sites
$(k-1,i)$ and $(k,j)$ on two adjacent circles labeled by $k-1$ and $k$, where $k=1,2,\dots,N_k$, will
be denoted by $Y^\mu_{k,j}(x_\mu)$. In analogy with
Eq.~(\ref{eq:links}), we expand this link field in the new
notation as
$Y^\mu_{k,j}=x^\mu+A^\mu_{k,j}(x_\mu)+\partial^\mu\phi_{k,j}(x_\mu)$.
From Eq.~(\ref{eq:warpedFP}), we then find for the total action
\begin{eqnarray}\label{eq:local}
\mathcal{S}_{\rm lin}&=&\mathcal{S}_{\rm FP}+
\int {\rm d}^4x\:M_4^2\Big[h^{0,1}_{\mu\nu}\Box h^{0,1}_{\mu\nu}+
\sum_{k=1}^{k_{\rm max}}\sum_{i=1}^{N_{k-1}}\Big(\epsilon^{2k}h^{k,i}_{\mu\nu}\Box
h^{k,i}_{\mu\nu}+m_*^2\epsilon^{4k}
(h^{k-1,i}_{\mu\nu}-h^{k,n_i}_{\mu\nu})\Box\phi_{k,n_i}\nonumber\\
&&\qquad+m_*^2\epsilon^{4k}(h^{k-1,i}_{\mu\nu}-h^{k,n_i+1}_{\mu\nu})\Box\phi_{k,n_i+1}+m_\ast^2\epsilon^{4k}[(\Box\phi_{k,n_i})^3
+(\Box\phi_{k,n_i+1})^3]\Big)\Big],
\end{eqnarray}
where $n_i=2(i-1)+1$. In the basis of canonically normalized fields
$\hat{h}_{\mu\nu}^{k,i}=M_4\epsilon^kh_{\mu\nu}^{k,i}$, the action
$\mathcal{S}_\textrm{lin}$ reads
\begin{eqnarray}\label{eq:localcanonical}
\mathcal{S}_{\rm lin}&=&\mathcal{S}_{\rm FP}+
\int {\rm d}^4x\:\Big[\hat{h}^{0,1}_{\mu\nu}\Box\hat{h}^{0,1}_{\mu\nu}+
\sum_{k=1}^{k_{\rm max}}\sum_{i=1}^{N_{k-1}}\Big(\hat{h}^{k,i}_{\mu\nu}\Box
\hat{h}^{k,i}_{\mu\nu}+M_4m_*^2\epsilon^{3k}
(\epsilon\hat{h}^{k-1,i}_{\mu\nu}-\hat{h}^{k,n_i}_{\mu\nu})\Box\phi_{k,n_i}\nonumber\\
&+&M_4m_*^2\epsilon^{3k}(\epsilon\hat{h}^{k-1,i}_{\mu\nu}-\hat{h}^{k,n_i+1}_{\mu\nu})\Box\phi_{k,n_i+1}+M_4^2m_\ast^2\epsilon^{4k}[(\Box\phi_{k,n_i})^3
+(\Box\phi_{k,n_i+1})^3]\Big)\Big].
\end{eqnarray}
Like in the rough, local flat space approximation of
Sec.~\ref{sec:subgraph}, let us now set in Eq.~(\ref{eq:localcanonical})
$\epsilon=1$ but keep the higher powers $\epsilon^{3k}$ since $k$
can be large. Consider now the subgraph in Fig.~\ref{fig:tree} for the case
$k_\textrm{max}=2$. For this
subgraph, the kinetic mixing terms in Eq.~(\ref{eq:localcanonical}) can be
organized in a kinetic matrix that is proportional to
\begin{equation}\label{eq:kinetic}
\left(
\begin{matrix}
-1 &  -1  & 0 & 0 & 0 & 0 \\
1 &   0  & -\epsilon^3 & -\epsilon^3 & 0 & 0 \\
0 &   1  & 0 & 0  & -\epsilon^3 & -\epsilon^3 \\
0 & 0  & \epsilon^3 & 0  & 0 & 0\\
0 & 0  & 0 & \epsilon^3 & 0 & 0 \\
0 & 0  & 0 & 0 &  \epsilon^3 & 0 \\
0 & 0  & 0 & 0 & 0 & \epsilon^3
\end{matrix}
\right),
\end{equation}
where we can choose a labeling such that the the rows and columns of the kinetic matrix are spanned by
$(\hat{h}_{\mu\nu}^{0,1},\hat{h}_{\mu\nu}^{1,1},\hat{h}^{1,2}_{\mu\nu},\hat{h}^{2,1}_{\mu\nu},
\hat{h}^{2,2}_{\mu\nu},\hat{h}^{2,3}_{\mu\nu},\hat{h}^{2,4}_{\mu\nu})$ and
$(\phi_{1,1},\phi_{1,2},\phi_{2,1},\phi_{2,2},\phi_{2,3},\phi_{2,4})$,
respectively.
Since the top-left $3\times 2$ sub-matrix in Eq.~(\ref{eq:kinetic})
has the rank 2, we see that
the mixing between two scalar Goldstones $\phi_{k,i}$ and
$\phi_{l,j}$ is small and only of the order $\sim\epsilon^3$ for $k\neq l$,
{\it i.e.}, the Goldstones associated with different concentric circles mix only
little with each other. Let us expand the Goldstones as $\phi_{1,i}=\frac{1}{\sqrt{2}}\sum_{n=1}^2
e^{{\rm i}2\pi i\cdot n/2}\Phi_{1,n}$, for $i=1,2$, and
$\phi_{2,i}=\frac{1}{\sqrt{2}}\sum_{n=1}^2e^{{\rm i}2\pi i\cdot
n/4}\Phi_{2,n}$, for $i=1,2,3,4$. Rotating to the basis of graviton mass
eigenstates
$(\hat{H}_{\mu\nu}^{0,1},\hat{H}_{\mu\nu}^{1,1},\hat{H}_{\mu\nu}^{1,2},
\hat{H}_{\mu\nu}^{2,1},\hat{H}_{\mu\nu}^{2,2},\hat{H}_{\mu\nu}^{2,3},\hat{H}_{\mu\nu}^{2,4})$, as given in Eqs.~(\ref{eq:subgraphstates}), and going to the basis of Goldstones
$(\Phi_{1,1},\Phi_{1,2},\Phi_{2,1},\Phi_{2,2},\Phi_{2,3},\Phi_{2,4})$,
the kinetic matrix in Eq.~(\ref{eq:kinetic}) becomes
\begin{equation}
\left(
\begin{matrix}
0 &   0  & 0 & 0 & 0 & 0 \\
1 &   0  &   e^{{\rm i}\pi/4}\epsilon^3 & 0 & e^{-{\rm
i}\pi/4}\epsilon^3 & 0 \\
0 &   \sqrt{3}  & 0 & 0  & 0 & -\sqrt{\frac{2}{3}}\epsilon^3 \\
0 & 0  & \epsilon^3 & 0  & 0 & 0\\
0 & 0  & 0 & \epsilon^3 & 0 & 0 \\
0 & 0  & 0 & 0 &  \epsilon^3 & 0 \\
0 & 0  & 0 & 0 & 0 & \sqrt{\frac{7}{3}}\epsilon^3
\end{matrix}
\right),
\end{equation}
which is, up to corrections of the order $\sim\epsilon^3$, on diagonal form.
In a similar way, for general $k_\textrm{max}\geq 2$, the
kinetic mixing terms in Eq.~(\ref{eq:localcanonical}) are
approximately diagonalized  by transforming to the graviton mass
eigenbasis spanned by the fields $\hat{H}_{\mu\nu}^{k,j}$, as defined in
Sec.~\ref{sec:generalcase}, and by expanding the scalar Goldstones as
$\phi_{k,i}=\frac{1}{\sqrt{N_k}}\sum_{n=1}^{N_k}e^{{\rm i}2\pi
i\cdot n/N_k}\Phi_{k,n}$, where $k=1,2,\dots, k_{\rm max}$ and
$i=1,2,\dots,N_k$. In this basis, the action in
Eq.~(\ref{eq:localcanonical}) is then approximately
\begin{eqnarray}\label{eq:diagonal}
\mathcal{S}_{\rm ref}&=&\mathcal{S}_{\rm FP}+
\int {\rm d}^4x\:\Big[\hat{H}^{0,1}_{\mu\nu}\Box \hat{H}^{0,1}_{\mu\nu}+
\sum_{k=1}^{k_{\rm max}}\sum_{n=1}^{N_k}\Big(\hat{H}^{k,n}_{\mu\nu}\Box
\hat{H}^{k,-n}_{\mu\nu}\nonumber\\
&+&M_4m_\ast^2\epsilon^{3k}\hat{H}^{k,n}_{\mu\nu}\Box\Phi_{k,-n}
+\sum_{m=1}^{N_k}M_4^2\frac{m_\ast^2\epsilon^{4k}}{\sqrt{N_k}}
(\Box\Phi_{k,n})(\Box\Phi_{k,m})(\Box\Phi_{k,-n-m})\Big)\Big],
\end{eqnarray}
where we have used the definitions
$\hat{H}^{k,-n}_{\mu\nu}=\hat{H}^{k,N_k-n}_{\mu\nu}$ and
$\Phi_{k,-n}=\Phi_{k,N_k-n}$. From Eq.~(\ref{eq:diagonal}), we read
off the canonically normalized fields $\hat{\Phi}_{k,n}=M_4m_*^2
\epsilon^{3k}\Phi_{k,n}$. In canonical normalization, the tri-linear derivative
coupling terms in Eq.~(\ref{eq:diagonal}) become
\begin{equation}\label{eq:trilinear}
M_4^2\frac{m_\ast^2\epsilon^{4k}}{\sqrt{N_k}}
(\Box\Phi_{k,n})(\Box\Phi_{k,m})(\Box\Phi_{k,-n-m})
\rightarrow
\frac{1}{\sqrt{N_k}M_k m_*^4\epsilon^{4k}}
(\Box\hat{\Phi}_{k,n})(\Box\hat{\Phi}_{k,m})(\Box\hat{\Phi}_{k,-n-m}),
\end{equation}
where $M_k=M_4\epsilon^k$. Similar to
Sec.~\ref{sec:effectivetheory}, we then find that the modes
$\hat{\Phi}_{k,n}$ in Eq.~(\ref{eq:trilinear}) become strongly coupled
at a scale
\begin{equation}\label{eq:Lambdak}
\Lambda_k^{6D}=(\sqrt{N_k}M_k m_\ast^4\epsilon^{4k})^{1/5}.
\end{equation}
The important point is here the presence of the factor $N_k$, which
can render the modes $\hat{\Phi}_{k,n}$ more weakly coupled when
$N_k$ becomes exponentially large. Such a factor $N_k$ is
absent in 5D lattice gravity models, {\it i.e.}, the presence of $N_k$
in Eq.~(\ref{eq:Lambdak}) is a result of working in a 6D
setup. Setting, {\it e.g.}, the inverse radial lattice spacing equal to
$m_\ast=M_4$, it follows that $\Lambda^{6D}_k=N_k^{1/10}M_k$, which
would be a factor $N_k^{1/10}$ above the local Planck scale $M_k$. But
having $\Lambda_k^{6D}$ significantly above $M_k$ requires many sites
$N_k$ on the $k$th circle. In our special example in
Sec.~\ref{sec:subgraph}, with $m_\ast=M_4\approx M_\textrm{Pl}$, $\epsilon=(0.1)^{1/20}$,
$\ell =2^{1/20}$, $w\simeq (0.1)\times M_\textrm{Pl}$, and $k_\textrm{max}\simeq\mathcal{O}(10^2)$, we have
$\Lambda^{6D}_{k_\textrm{max}}\approx 3\times M_k$.  For the same
parameters, only with $\ell$ changed to $\ell\simeq 5^{1/20}$, we
obtain $N_{k_\textrm{max}}^{1/10}\simeq 10$ while keeping
$(\epsilon\cdot\ell)^{20}\simeq 0.5$ sufficiently small. Also, in our
example in Sec.~\ref{sec:subgraph}, where $m_\ast\simeq (0.1)\times
M_\textrm{Pl}$, $\epsilon\simeq 0.1$, and $w\simeq (0.2)\times
M_\textrm{Pl}$, we obtain for $\ell=4$ and $k_\textrm{max}=15$ a
strong coupling scale $\Lambda^{6D}_{k_\textrm{max}}\approx (1.3)\times M_{k_\textrm{max}}$.

The scale $\Lambda^{6D}_k$ in Eq.~(\ref{eq:Lambdak}) has been found by first going in
Eq.~(\ref{eq:localcanonical}) to the rough, local flat space
approximation and then neglecting the mixing between the modes on neighboring
circles. We will now, however, be interested in taking the nonzero
mixing effects between the modes on neighboring circles into
account. In doing so, we will follow closely the argumentation presented for the warped 5D
case in Ref.~\cite{Randall:2005me}, and for the rest of this section,
we will restrict ourselves to the interesting case where $m_*=M_4\approx
M_\textrm{Pl}$ and $w\simeq (0.1)\times M_\textrm{Pl}$. Like in the 5D example, the modes
which are relevant for the strong coupling scale at some site have a maximum
wavelength in radial direction which is of the order of the inverse curvature
scale $1/w$, and it is these long-wavelength modes which become most
earliest strongly coupled. Considering within distances $\sim 1/w$ in
radial direction the space as locally flat and treating the modes in
this regime as plane waves, it is seen in our
calculation that the mixing of the
modes between neighboring circles can be approximately taken into
account by sending in Eq.~(\ref{eq:Lambdak})
$m_\ast^4\epsilon^{4k}\rightarrow (w/M_4)^4m_\ast^4\epsilon^{4k}$ and
$\sqrt{N_k}\rightarrow (M_4/w)^{3/2}\sqrt{N_k}$. We thus obtain from
Eq.~(\ref{eq:Lambdak}) a local strong coupling scale at the $k$th
circle that is approximately
\begin{equation}\label{eq:localscale}
\Lambda_\textrm{warp}^{6D}=\sqrt{\frac{w}{M_4}}(\sqrt{N_{k}}M_k m_\ast^4\epsilon^{4k})^{1/5}.
\end{equation}
With respect to $\Lambda^{6D}_{k}$ in Eq.~(\ref{eq:Lambdak}), the
local strong coupling scale $\Lambda^{6D}_\textrm{warp}$ is lowered by
a factor $(w/M_4)^{1/2}$, which is, for our choice $w\simeq (0.1)\times M_4$, of the order $(w/M_4)^{1/2}\approx
0.3$. Setting in Eq.~(\ref{eq:localscale})
$m_\ast=M_4$, one arrives at
$\Lambda^{6D}_\textrm{warp}=(w/M_4)^{1/2}N_k^{1/10}M_k$. In our
example in Sec.~\ref{sec:subgraph}, with $m_\ast=M_4\approx M_\textrm{Pl}$, $w\simeq(0.1)\times M_\textrm{Pl}$, $\epsilon=(0.1)^{1/20}$, and
$\ell=2^{1/20}$, the local strong coupling scale will thus be pushed
to values higher than the local Planck scale $M_k$ when $k$ is in the
regime $k\gtrsim\mathcal{O}(10^2)$. For $\ell=3^{1/20}$, {\it e.g.}, this
would, of course, happen for smaller $k$. Note also that $\Lambda^{6D}_\textrm{warp}$ reproduces
in the limit $N_k\rightarrow 1$ the local strong coupling
scale in 5D warped space given in Ref.~\cite{Randall:2005me}.

It is instructive to compare $\Lambda_\textrm{warp}^{6D}$ with the strong
coupling scales in 5D lattice gravity. In 5D flat space, the strong
coupling scale is $\Lambda^{5D}_\textrm{flat}=(N^{3/2}M_4m^4_1)^{1/5}$,
where $m_1$ is the mass of the lightest graviton and $N$ is the number of lattice
sites \cite{Arkani-Hamed:2003vb}. Since $m_1=m/N$, where $m$ is the inverse lattice spacing, we
see that, even for $m$ as large as $M_4$, the strong coupling scale
$\Lambda^{5D}_\textrm{flat}$ is always smaller than the local Planck
scale $M_4$ and that $\Lambda^{5D}_\textrm{flat}$ goes to zero in the
large volume limit $N\rightarrow \infty$. This situation is remedied
in 5D warped space, where the local strong coupling scale is
$\Lambda^{5D}_\textrm{warp}=(w/M_4)^{1/2}M_k$, in which $w$ is the 5D
curvature scale and $M_k=M_4e^{-kw/m}$ is the local Planck scale at the $k$th site \cite{Randall:2005me}. The strong coupling scale
$\Lambda^{5D}_\textrm{warp}$ is independent from $N$, which allows to
take the large volume limit like in RS II \cite{Randall:1999vf}. However,
$\Lambda^{5D}_\textrm{warp}$ is still smaller by a factor
$(w/M_4)^{1/2}$ than $M_k$. On our warped hyperbolic disk, on the
other hand, $\Lambda^{6D}_\textrm{warp}$ is (i) independent from the
number of concentric circles $k_\textrm{max}$ and can (ii) become as large
as the local Planck scale $M_k$ when approaching the boundary. It is,
however, important to keep in mind that the local strong coupling scale
$\Lambda^{6D}_\text{warp}$ in Eq.~(\ref{eq:localscale}) has been
calculated by restricting to the Goldstone boson sector without taking
the coupling to matter into account. Next, in
Sec.~\ref{sec:finegrainedmatter}, we will analyze the strong coupling
scale that is actually seen by an observer made of SM matter localized
at a single site on the boundary of the disk.

\subsection{Strong coupling scale for an observer}\label{sec:finegrainedmatter}
To determine the strong coupling scale that an observer would actually
see, we have to take the coupling of the Goldstone boson fields to
matter into account. By the same arguments as in
Sec.~\ref{sec:coarsematter}, we find from $\Lambda^{6D}_\text{warp}$ in Eq.~(\ref{eq:localscale}) that the
local strong coupling scale $\Lambda^{6D}_\text{bound}$ that is seen
by an observer localized on a single site on the boundary of the
fine-grained model is
\begin{equation}\label{eq:Lambda6Dmatter}
\Lambda^{6D}_\text{bound}=\sqrt{\frac{w}{M_4}}(M_{k}
m_\ast^4\epsilon^{4k})^{1/5},
\end{equation}
where $k=k_\text{max}$. In the limit $m_\ast\rightarrow M_4$, we have
$\Lambda^{6D}_\text{bound}\rightarrow(w/M_4)^{1/2}M_k$, which is the local
strong coupling scale in discretized 5D warped space with curvature
scale $w$ that is observed at a site with local Planck scale
$M_k$. This space has the same background geometry as the 5D
sub-manifold of the disk that is given by the geodesic
line connecting the center with the boundary on the disk. One might then wonder under what
circumstances the fine-grained model can improve the strong coupling
behavior of 5D lattice gravity.

\begin{figure}
\begin{center}\includegraphics*[bb = 180 625 510 740]{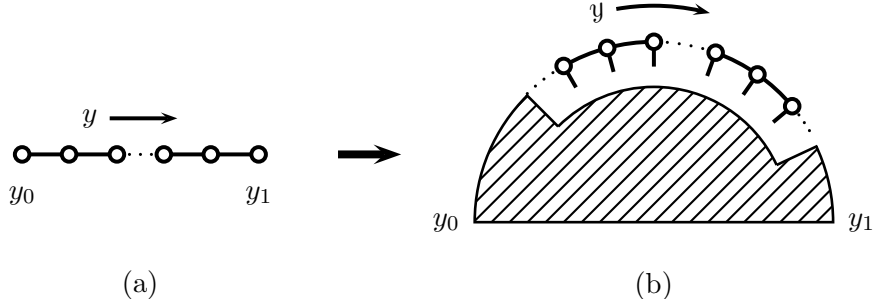}
\end{center}
\vspace*{-5mm}
\caption{{\small Comparison of a discretized gravitational 5th
   dimension (a) with the discrete 5D subspace on the boundary of the
   hyperbolic disk (b).
   Figure (b) shows the graph (a) wrapping the hyperbolic
   disk as a boundary of the fine-grained model (depicted is only a part
   of the discrete boundary, and the hatched region represents the remaining
   sites and links of the discretized disk).}}
\label{fig:wrapped}
\end{figure}

To address this question, we can view the boundary of the fine-grained
model as a graph of a discretized 5th
dimension that is ``wrapped'' around the disk as shown
in Fig.~\ref{fig:wrapped}. The interval (a) and the boundary of the
disk (b) in Fig.~\ref{fig:wrapped} have identical size and are
described by the same 5D background geometry with common coordinate $y$: they have the same number of lattice sites, proper inverse
lattice spacings, and local Planck scales. Similar to what we saw in
the discussion of the coarse-grained model in
Sec.~\ref{sec:coarsegrained}, the local strong coupling scale $\Lambda^{6D}_\text{bound}$ on the 5D boundary in Fig.~\ref{fig:wrapped} (b) can be much larger than the local strong coupling scale of the
unwrapped discretized 5D interval in Fig.~\ref{fig:wrapped} (a).
This is a result of the fact that the boundary sites in graph (b) are
-- different from the sites of the graph (a) -- also connected by
radial links to the interior of the disk. Let us investigate this more explicitly for the case
where the 5D spaces in Fig.~\ref{fig:wrapped} exhibit a non-trivial
warp-factor in $y$ direction.

The geometry of the hyperbolic disk that we are interested in has both a
radial dependence of the warp factor as well as a non-zero warping in $\varphi$ direction. In other words, in the language of
Sec.~\ref{sec:continuum}, the metric is as in Eq.~(\ref{eq:6D-metric})
with a new warp factor $e^{2\sigma(r,\varphi)}$, where
\begin{equation}\label{eq:angularwarping}
\sigma(r,\varphi)=-w\cdot r\big(1-\frac{\varphi}{\pi}\big).
\end{equation}
As before, the proper radius of the hyperbolic disk is $L$. Different
from earlier, we have now a single IR brane at $(r,\varphi)=(L,0)$
and a set of UV
branes that are located on a straight line extending in $\varphi=\pi$
direction from the center to
the point $(L,\pi)$ on the boundary. The SM with the observer is
assumed to reside on the boundary at the IR brane.

We define the common 5D geometry of the interval and the boundary in Figs.~\ref{fig:wrapped}
(a) and (b) by restricting the coordinates on the
disk to the 5D boundary sub-manifold according to $(r,\varphi)\rightarrow
(L,\varphi)$. The 5D background metric is given by
\begin{equation}
\textrm{d}s^{2}=e^{2\sigma(y)}g_{\mu\nu}(x)\textrm{d}x^{\mu}\textrm{d}x^{\nu}-\textrm{d}y^2,\label{eq:5D-metric}
\end{equation}
where $\sigma(y)=-wvLy/[\pi\,\text{sinh}(vL)]$ and $y$ is the 5D
coordinate. In terms of the disk parameters, we have
$y=y(\varphi)=(\pi-\varphi)v^{-1}\text{sinh}(vL)$, {\it i.e.}, $y$ is equal
to the proper distance measured along the boundary between the point
$(L,\varphi)$ and the UV brane at $(L,\pi)$. In
Figs.~\ref{fig:wrapped} (a) and (b) it is $y_0=y(\pi)$ and $y_1=y(0)$.

We discretize the hyperbolic disk and
use the same notation exactly as described in
Secs.~\ref{sec:structure} and \ref{sec:subgraph}. In the  discretized
model, the warping in $\varphi$ direction is implemented by redefining
in Eqs.~(\ref{eq:Swarp}) and (\ref{eq:warpedFP}) the quantity
$\epsilon^k$ as $\epsilon^k\rightarrow
\epsilon^{k}\epsilon^{-k\cdot i/N_k}$, where $i$ labels the $i$th site
on the $k$th circle. For example, the local Planck scale on a site
$(k,i)$ is now $M_{k,i}\equiv M_4\epsilon^{k(1-i/N_k)}$, and the
canonically normalized fields are
$\hat{h}^{k,i}_{\mu\nu}=M_4\epsilon^{k(1-i/N_k)}h^{k,i}_{\mu\nu}$.
In the rough, local flat space approximation of
Sec.~\ref{sec:subgraph}, we then obtain for the mass matrix element between
two fields $\hat{h}^{k,i}_{\mu\nu}$ and
$\hat{h}^{l,j}_{\mu\nu}$ in the limit $N_k\gg 1$ the expression
\begin{equation}\label{eq:fineapproximation}
M_{(k,i)(l,j)}\approx\frac{1}{a_r^2}\epsilon^{2(l+1)(1-i/N_k)}[(2+\epsilon^{-2})\delta_{k,l}\delta_{i,j}-
\delta_{k+1,l}(\delta_{n_i,j}+\delta_{n_i+1,j})-\delta_{k-1,l}\delta_{\lceil
i/2\rceil,j}],
\end{equation}
where $k>0$. Again, for $\epsilon\cdot \ell\ll 1$ ({\it e.g.},
$\epsilon\cdot \ell\lesssim 0.5$), the usual
4D Planck scale $M_\text{Pl}\approx 10^{18}\text{GeV}$ is set by the
local Planck scale $M_4$ on the UV branes, and we have
$M_\text{Pl}\approx M_4$. In general, since the UV branes are spread
out on the disk around the straight line connecting the center with
the point $(L,\pi)$ on the boundary, the fundamental scale
$M_4$ will be reduced a little by a volume suppression factor; but this effect is small and can be
neglected in the regime $k_\text{max}\simeq 10-10^2$,
where $M_4$ becomes suppressed by only one or two orders of magnitude.

Switching on a non-trivial warp factor in $\varphi$ direction will
practically not change $\Lambda^{6D}_\text{bound}$ in
Eq.~(\ref{eq:Lambda6Dmatter}), since, for a large hyperbolic
curvature of the disk, the strong coupling scale on the boundary is
virtually insensitive to the circumference of the disk. Let us next
compare the observed strong coupling scales for the two graphs (a)
and (b) in Fig.~\ref{fig:wrapped}: Graph (b)
exhibits at the IR brane the local strong coupling scale
$\Lambda^{6D}_\text{bound}$ given in Eq.~(\ref{eq:Lambda6Dmatter}). In
contrast to this, we find after discretization of the 5D background in Eq.~(\ref{eq:5D-metric})
that the strong coupling scale for an observer localized on the
IR brane of the discretized interval in
graph (a) at $y=y_1$ is given by
\begin{equation}
\Lambda^{5D}_\text{warp}=(wvL)^{1/2}[\pi M_4\,\text{sinh}(vL)]^{-1/2}(M_km_\ast^4\epsilon^{4k})^{1/5},\label{eq:Lambda5D}
\end{equation}
where $k=k_\text{max}$. Thus, for large $vL$, the strong coupling scale $\Lambda^{5D}_\text{warp}$ in
Eq.~(\ref{eq:Lambda5D}) drops below
$\Lambda^{6D}_\text{bound}$ in Eq.~(\ref{eq:Lambda6Dmatter}). This means that the observed local strong coupling scale of
5D lattice gravity can be increased by wrapping the discretized 5th dimension around a hyperbolic disk. This requires, however, that the warping in $y$
direction is not too large: If $vL\sim 1$, then
$\Lambda^{6D}_\text{bound}$ and $\Lambda^{5D}_\text{warp}$ in
Eq.~(\ref{eq:Lambda5D}) will become identical.

Let us estimate the 5D curvature scales $-\sigma(y)/y$ for which
$\Lambda^{6D}_\text{bound}$ becomes significantly larger than
$\Lambda^{5D}_\text{warp}$ in Eq.~(\ref{eq:Lambda5D}). We consider the specific
example given at the end of Sec.~\ref{sec:subgraph}, where we formally take
$\epsilon=(0.1)^{1/20}$ and $\ell=\ell^{1/20}$. Here, the proper radius is
$k_\text{max}\simeq\mathcal{O}(10^2)$ units $M_4^{-1}$ wide and
$m_\ast\simeq M_4$. The 6D curvature scale is $w=M_\text{Pl}/10$ and $M_4\approx M_\text{Pl}$. On the boundary, the number of sites is
$N_{k_\text{max}}\simeq \mathcal{O}(10^4)$, and the local Planck scale
at the IR brane is $M_{k_\text{max}}=\epsilon^{k_\text{max}}M_\text{Pl}\simeq
1\,\text{TeV}$. The curvature scale for the warping in $y$ direction
is $-\sigma(y)/y\simeq 10^{-3}\times w$, and the strong coupling scale
$\Lambda^{5D}_\text{warp}$ in Eq.~(\ref{eq:Lambda5D}) is by a factor
$\sim 100$ smaller than $\Lambda^{6D}_\text{bound}$ in
Eq.~(\ref{eq:Lambda6Dmatter}). We thus find that the strong coupling scale
seen by an observer located on the 5D boundary sub-manifold of the disk can be
significantly larger than the strong coupling scale in the
corresponding discretized 5D warped space for values
$-\sigma(y)/y\lesssim M_4\times 10^{-3}$ of the 5D curvature scale.

\section{Application to Dirac neutrino masses}\label{sec:neutrinomasses}
So far, we have only been considering the case where gravity is
propagating on the disk. In this section, we will be concerned with
matter in the bulk and formulate a model for small Dirac
neutrino masses by implementing on the hyperbolic disk a discretized
version of the volume suppression mechanism of
Ref.~\cite{Arkani-Hamed:1998vp}. The volume suppression mechanism in the large extra dimensional
scenario is attractive, since it yields a purely geometric origin of
neutrino masses. For simplicity, we shall restrict here
to the coarse-grained model described in Secs.~\ref{sec:coarsegrained}
and \ref{sec:effectivetheory}.
Note that, in the following, gravity is not really relevant for generating
neutrino masses, and we could equally well have non-gravitational extra
dimensions and couple matter only to a single copy of gravity. We work, nevertheless, in the theory space of
Sec.~\ref{sec:coarsegrained}, since it allows to avoid the
experimental bounds on extra KK-neutrinos and KK-gravitons for the same reason: The curvature of the
disk generates a big mass gap between the zero mode and the higher
KK-excitations that is not related to the large bulk volume.

The parent continuum theory which we discretize contains a massless
right-handed (RH) SM singlet bulk neutrino that can propagate on the
disk, or some sub-space thereof, while the SM fields are all located on a single
point on the disk.
Specifically, we consider
the total action of the bulk neutrino on the disk as a combination of
the following two limiting cases: (i) the bulk neutrino propagates only
within the star sub-geometry of the disk and (ii) the bulk neutrino
propagates only in the circle sub-geometry.

Let us first consider case (i): the action for the gravitons is
described by $\mathcal{S}_{\rm disk}$ in Sec.~\ref{sec:disk}, but the
latticized RH bulk neutrino propagates only in the star
sub-geometry. The bulk neutrino is
represented in the discretized theory by putting on each site
$i$ one 4D RH SM singlet Dirac fermion
$\Psi_i=(\nu_{iR},\overline{\nu^c_{iR}})$, where $\nu_{iR}$ and
$\nu_{iR}^c$ are 4D two-component Weyl spinors and  $i=0,1,\dots, N$.
The star sub-geometry, with $N$ sites surrounding the center, can be
thought as being
composed of $N$ two-site models that are glued together at the site
$i=0$. It is convenient to label these two-site models as $(0,i)$, where
$i=1,2,\dots ,N$. To obtain here the action of the RH bulk
neutrino, we consider each two-site model $(0,i)$ as the coarse-grained limit of a latticized fifth dimension that is compactified on an interval $[0,R_i]$ and apply discretized orbifold boundary conditions at the endpoints. In this picture, the continuum limit of
the star sub-geometry is a configuration of $N$ continuous intervals
$[0,R_i]$ which intersect in their common endpoint 0. The RH
bulk neutrino propagates in this continuum theory as a 5D field on the
intervals and is subject to orbifold boundary conditions at the
endpoints of the segments $[0,R_i]$. We denote the 5D bulk neutrino by
$\Psi=(\nu_R,\overline{\nu^c_R})$, where $\nu_R$ and $\nu^c_R$ are 5D
two-component Weyl spinors, and assume on each of the $N$
intervals the Neumann and Dirichlet boundary conditions
$\left .(\partial \nu_R/\partial y_i)\right|_{y_i=0,R_i}=0$ and
$\left. \nu^c_R\right|_{y_i=0,R_i}=0$,
where $y_i\in [0,R_i]$ is the coordinate of the interval $[0,R_i]$ and
$i=1,2,\dots ,N$. The discretization of these boundary conditions
has been described in Refs.~\cite{Hill:2000mu,Bhattacharya:2005xa,Skiba:2002nx}
and produces, in the limit where each interval $[0,R_i]$ is replaced by
a two-site model $(0,i)$, from the latticized kinetic term of $\Psi$
the neutrino mass terms
\begin{equation}
\mathcal{L}_{(0,i)}=m_\ast(\nu_{iR}\nu^c_{iR}-\nu_{0R}\nu^c_{iR})+{\rm
h.c.},
\end{equation}
where we have chosen a common length $R_i\equiv R
\sim m_\ast^{-1}$ for all intervals. In the flat limit
$g_{\mu\nu}\rightarrow \eta_{\mu\nu}$, we then take the total action
for the bulk neutrino in the star sub-geometry $\mathcal{S}_{\rm
star}^\Psi$ to be
$\mathcal{S}_{\rm star}^\Psi=\int {\rm d}^4x(\sum_{i=0}^N
{\rm i}\overline{\Psi}_i\cancel{\partial}\Psi_i
+\sum_{i=1}^N\mathcal{L}_{(0,i)})$.
To include the effect from the boundary, let us next
consider case (ii): the latticized RH bulk neutrino
propagates only on the boundary of the disk. For this case, we take the action of the latticized RH
neutrino on the circle $\mathcal{S}_{\rm circle}^\Psi$ to be of the Wilson-Dirac form
$\mathcal{S}_{\rm circle}^\Psi  =
\int\textrm{d}^{4}x\sum_{i=1}^N({\rm i}\overline{\Psi}_{i}\,
\cancel{\partial}\,\Psi_{i}+
\mathcal{L}_{(i,i+1)})$, where we have introduced a discretized kinetic term
\begin{equation}
\mathcal{L}_{(i,i+1)}=m\cdot\nu_{iR}(\nu^c_{(i+1)R}-\nu^c_{iR})+{\rm h.c.},
\end{equation}
for each pair of sites $(i,i+1)$ on the boundary of the disk. For definiteness, we suppose also that $N$ is even. The action
$\mathcal{S}_{\rm circle}^\Psi$ has been widely discussed in the
literature as a standard example for a fermion propagating in a
latticized fifth dimension compactified on the circle $S^1$
\cite{Hill:2000mu} (see also, {\it e.g.},
Refs.~\cite{Bauer:2003mh,Hallgren:2004mw}). Now, we define the
total action of the bulk neutrino on the disk
$\mathcal{S}_{\rm disk}^\Psi$ by simply adding
the mass terms $\mathcal{L}_{(i,i+1)}$ to $\mathcal{S}_{\rm
star}^\Psi$ such that
\begin{equation}\label{eq:diskneutrino}
\mathcal{S}_{\rm disk}^\Psi=\int {\rm d}^4x\big{(}
\sum_{i=0}^N\overline{\textrm{i}\Psi}_i\cancel{\partial}\Psi_i+
\sum_{i=1}^N[\mathcal{L}_{(0,i)}+C\cdot\mathcal{L}_{(i,i+1)}]
\big{)},
\end{equation}
where $C$ is some suitable dimensionless parameter. The mass
terms in Eq.~(\ref{eq:diskneutrino}) give rise to the $(N+1)\times (N+1)$ Dirac neutrino mass matrix\begin{equation}\label{eq:neutrinomassmatrix1}
M_D=
m_\ast
\left(
\begin{matrix}
0 & -1 & -1 & -1 & \cdots &-1\\
0 & 1  &  0 &  0 & \cdots & 0\\
0 & 0  &  1 &  0 & \cdots & 0\\
0 & 0  &  0 &  1 & \ddots  &   \vdots\\
\vdots & \vdots & \vdots & \ddots &\ddots& 0\\
0 &0 & 0 & \cdots & 0 &1
\end{matrix}
\right)-
Cm\left(
\begin{matrix}
0&0&0&0&\cdots&0\\
0&-1 &  1 &  0 & \dots &0\\
0& 0  & -1 &  1  &\ddots & \vdots\\  
\vdots& \vdots  & 0 & \ddots &\ddots & 0\\
0& 0  & \vdots  &  \ddots     &  -1 & 1\\
0& 1  & 0 &  \dots     &   0  &-1\\                
\end{matrix}
\right),
\end{equation}
where the rows and columns are spanned by
$(\nu_{0R},\nu_{1R},\dots,\nu_{NR})$
and
$(\nu^c_{0R},\nu^c_{1R},\dots,\nu^c_{NR})$,
respectively.
We can choose the parameter $C$ such that the matrix
$M_DM_D^\dagger$ becomes identical with the graviton mass matrix
$M_g^2$ in Eq.~(\ref{eq:Mg}), in which case the RH neutrino masses are
given by Eq.~(\ref{eq:eigenvalues}). The mass matrix $M_DM_D^\dagger$ is then diagonalized by expanding
\begin{subequations}\label{eq:neutrinostates}
\begin{eqnarray}
\nu_{0R}& = &
\frac{1}{\sqrt{N+1}}\hat{\nu}_{0R}-\frac{N}{\sqrt{N(N+1)}}
\hat{\nu}_{NR},\\
\nu_{iR} & = & \frac{1}{\sqrt{N}}\sum_{n=1}^{N-1}e^{-{\rm i}2\pi
i\cdot n/N}\hat{\nu}_{nR}+
\frac{1}{\sqrt{N+1}}\hat{\nu}_{0R}+\frac{1}{\sqrt{N(N+1)}}\hat{\nu}_{NR},
\end{eqnarray}
\end{subequations}
where $i=1,2,\dots ,N$ and $\hat{\nu}_{nR}$ ($n=0,1,\dots, N$)
is the $n$th mass eigenstate belonging to the mass eigenvalue $M_n^2$
given in Eq.~(\ref{eq:eigenvalues}). In Eqs.~(\ref{eq:neutrinostates}), we thus have one zero mode
neutrino $\hat{\nu}_{0R}$ with flat profile, one heavy
neutrino $\hat{\nu}_{NR}$ with mass $M_N\sim m_\ast\sqrt{N}$, and~$N-1$
states $\hat{\nu}_{1R},\hat{\nu}_{2R},\dots,\hat{\nu}_{N-1}$
with masses $M_n$ that become $M_n\approx m_\ast$, for $m_*\gg m$.

Let us now introduce the SM neutrinos by adding all SM fields on a
single site on the boundary, {\it e.g.}, on the site $i=1$.
To simplify the discussion, we assume like in
Ref.~\cite{Arkani-Hamed:1998vp} that $B-L$ is conserved in the bulk
and assign the bulk neutrino a
$B-L$ number opposite to that of the SM leptons. On the site $i=1$, we
then have a local Yukawa interaction $\mathcal{L}_{\rm int}=
f_\alpha\ell_{\alpha}\textrm{i}\sigma_2H\nu_{1R}+\textrm{h.c.}$, where $\ell_{\alpha}=(\nu_\alpha,\:e_\alpha)^T$ are the lepton
doublets with generation index $\alpha=1,2,3$, while $H$ is the Higgs
doublet, and $f_\alpha$ is a dimensionless
order one Yukawa coupling. Going to momentum basis,
$\mathcal{L}_\textrm{int}$ approximately reads
\begin{equation}\label{eq:wavefunctionsuppression}
\mathcal{L}_{\rm int}\approx
f_\alpha \frac{\langle H\rangle}{\sqrt{N}}\nu_{\alpha}\hat{\nu}_{0R}+\textrm{h.c.}
\end{equation}
and produces a Dirac
neutrino mass term between the active neutrinos
$\nu_\alpha$ and the zero mode $\hat{\nu}_{0R}$ that is
suppressed by a volume factor $\sqrt{N}=\sqrt{Rm}$. This is the analog
of the volume suppression mechanism in
Ref.~\cite{Arkani-Hamed:1998vp}. In the coarse-grained model, for
$m_\ast\sim m$ and $x\equiv v/m_\ast\gtrsim 1$, we have $N\simeq
\pi\,\textrm{exp}(x-\textrm{ln}\,x)$, {\it i.e.}, a moderately large
curvature $v$ [{\it e.g.}, $v/m_*=\mathcal{O}(10)$] gives an exponentially large number of sites $N$. For $\langle H\rangle\simeq 10^2\:{\rm GeV}$ and $N\simeq 10^{24}$ sites, the Dirac neutrino masses will be of the
right order $\sim 10^{-2}\:{\rm eV}$, which would correspond to
$v/m_\ast\simeq 60$ in the above approximation. The important point is
here that the strong coupling scale is, even in the large $N$
(or large volume) limit, always bounded from below by
$\Lambda_\textrm{disk}=(M_\textrm{Pl}m_\ast^4)^{1/5}$. Since all
massive neutrino singlets have a mass larger than
$m_\ast$, it follows that for $m_\ast\gtrsim 1\:{\rm GeV}$ all
constraints on KK neutrinos from astrophysics and the early universe
\cite{Barbieri:2000mg} are avoided.

Note that we have been using here the coarse-grained model with
many sites on the boundary as a tool for estimating quickly the
effect of putting a RH neutrino in the bulk with large volume. The
number of sites $N$ measures the size of the hyperbolic
disk in the corresponding continuum theory, where the same $N$
would be simply interpreted as the number of KK modes below the
fundamental Planck scale. Therefore,
Eq.~(\ref{eq:wavefunctionsuppression}) is exactly on the same footing
as the well-known continuum result of Ref.~\cite{Arkani-Hamed:1998vp}, and
provides an attractive origin for small Dirac neutrino masses.

\section{Summary and Conclusions}
In this paper, we have investigated gravity in a 6D
geometry, where the two extra dimensions are discretized and form a hyperbolic
disk with constant curvature. We have studied two types of discretizations of the disk.
In our first model, we have considered a coarse-grained discretization
with $N$ sites on the boundary and one site in the center
of the disk. For this case, we have determined the mass spectrum and
mass eigenstates of
the gravitons and found in the limit of large curvature a typical
KK-type spectrum sitting on top of a large mass gap between the zero mode
and the first massive mode. This mass gap is set by the inverse
radius of the disk while the other modes become essentially
degenerate in mass. This feature allows to avoid all existing constraints
on KK gravitons from Cavendish-type experiments, astrophysics, and
cosmology. Additionally, this model contains a single massive mode
that becomes very heavy in the large~$N$ limit. We have also discussed
collider signatures of the coarse-grained model at the LHC and at a possible
future linear collider.

The strong coupling scale in this coarse-grained model without couplings to matter  converges for large $N$ to a value like in the theory of a
single massive graviton, where the graviton has a mass of the order of
the inverse proper radius of the disk. This is completely different
from a discrete gravitational extra dimension in 5D flat space, which exhibits a UV/IR
connection problem, where the strong
coupling scale would eventually go to zero when taking the large $N$
limit. This UV/IR connection problem, however, is absent in the
coarse-grained model, where a sensible EFT is defined in the large $N$ limit. Even if interactions with matter are included, the theory remains valid, but with a lowered strong coupling scale. In this model, we have also studied an implementation of a bulk fermion that allows to
generate small Dirac neutrino masses in the limit of a large curvature
of the disk via a discrete version of the well-known
volume suppression mechanism for Dirac neutrino masses in flat extra
dimensions. Again, due to the particular form of the spectrum, all
experimental and observational constraints on the massive KK-type neutrinos are avoided.

In our second model, we describe a fine-grained discretization of the
hyperbolic disk with nonzero warping along in radial and angular direction. In
this setup, the sites are situated on equidistant concentric circles
and, as a result of the curvature of the disk, the number of sites on the circles grows exponentially in radial
direction. We have calculated the graviton mass eigenvalues and mass
eigenstates in the fine-grained model by going, as suggested
previously in an analysis of the discretized 5D RS model, to a rough, local flat
space approximation. Moreover, we have determined in the fine-grained
model a local strong coupling scale from the kinetic mixing matrix between the
gravitons and scalar Goldstones. The existence of a local
strong coupling scale in the fine-grained model is, like in the
corresponding 5D case, a result of the warping, which avoids the UV/IR
connection problem from flat space by locally introducing an effective
size or volume of the extra dimensions.
Taking the coupling to matter into account, we find that the local
strong coupling scale seen by a brane-localized observer on the
boundary can become as large as in a discretized 5D warped
model which has the background geometry of a geodesic line connecting
the center with the boundary. Moreover, it turns out that the observed
strong coupling behavior of gravity in discretized 5D warped space can be improved in six
dimensions by wrapping the graph of the fifth dimension as a boundary
around the fine-grained model. This holds for a 5D curvature scale
that is roughly at least by a factor $10^{-3}$ smaller than the fundamental scale.

In conclusion, we have seen for different implementations of two discrete
gravitational extra dimensions compactified on a hyperbolic disk that
a high curvature or strong warping of the disk allows to avoid the UV/IR connection problem of
lattice gravity in flat space. This observation is similar to the
result in 5D warped space. By going to six dimensions on the hyperbolic disk,
however, it is, for a range of 5D curvature scales, possible to
further improve on the boundary the strong coupling behavior of lattice gravity in 5D
warped space. Moreover, a strong curvature of the disk allows to avoid all constraints on KK-states
and can, {\it e.g.}, be employed for generating small
Dirac neutrino
masses in agreement with experiment.

It would be interesting to relate our analysis, {\it
e.g.},  also to moduli stabilization in
effective theories, to theories with
spontaneously broken space-time symmetries \cite{Kirsch}, and to studies
on the implications of latticized extra dimensions for cosmology
\cite{Bauer:2005hb}.

\section*{Acknowledgements}
We would like to thank K.S.~Babu, I.~Bengtsson, M.~Blennow,
T.~Enkhbat, N.~Kauer, T.~Konstandin,
T.~Ohlsson, and M.D.~Schwartz for useful comments and
discussions. This work was supported by the ``Sonderforschungsbereich 375 f\"ur
Astroteilchenphysik der Deutschen Forschungsgemeinschaft'' (F.B.), the G\"oran Gustafsson
Foundation (T.H.), and the U.S.~Department of Energy under grant
number DE-FG02-04ER46140 (G.S.).

\appendix

\section{\label{sec:Action-disk} Gravitational action on the hyperbolic disk}
In this section, we determine the gravitational action on the
hyperbolic disk $K_2$ introduced in Sec.~\ref{sec:continuum}.
We start with the 6D metric $\tilde{g}_{MN}$ as defined in
Eq.~(\ref{eq:6D-metric}), where $\tilde{g}_{\mu\nu}=
e^{2\sigma(r)}g_{\mu\nu}$, $\tilde{g}_{55}=-1$, and
$\tilde{g}_{66}=-\textrm{sinh}^2(vr)/(v^2)$. Partial and covariant
derivatives are denoted by commas and semicolons, respectively. In these
coordinates, the nonzero Christoffel symbols are given by
\begin{eqnarray}
\Gamma_{\mu\nu}^{\sigma} & = &
\frac{1}{2}\tilde{g}^{\sigma\rho}(\tilde{g}_{\mu\rho,\nu}+\tilde{g}_{\rho\nu,\mu}-\tilde{g}_{\mu\nu,\rho}),\nonumber\\
\Gamma_{\nu5}^{\mu}&=&\frac{1}{2}\tilde{g}^{\mu\rho}\tilde{g}_{\nu\rho,5},\quad\Gamma_{\nu6}^{\mu}\:=\:\frac{1}{2}\tilde{g}^{\mu\rho}\tilde{g}_{\nu\rho,6},\quad
\Gamma_{\mu\nu}^{5}\:=\:-\frac{1}{2}\tilde{g}^{55}\tilde{g}_{\mu\nu,5},\quad
\Gamma_{\mu\nu}^{6}\:=\:-\frac{1}{2}\tilde{g}^{66}\tilde{g}_{\mu\nu,6},\nonumber\\
\Gamma_{66}^{5} & = &
-\frac{1}{2}\tilde{g}^{55}\tilde{g}_{66,5}\:=\: - \frac{1}{2v}\,
\textrm{sinh}\,(2vr),\quad\Gamma_{56}^{6}\:=\:
\frac{1}{2}\tilde{g}^{66}\tilde{g}_{66,5}\:=\:v\,\textrm{coth}(vr),\end{eqnarray}
where $\Gamma^A_{BC}=\Gamma^A_{CB}$. Note that, due to the block-diagonal form of the metric
$\tilde{g}_{MN}$, the internal summation in $\Gamma^\sigma_{\mu\nu}$
runs only over 4D indices. The 6D Ricci tensor is defined by
$\tilde{R}_{MN}=\Gamma_{MA,N}^{A}-\Gamma_{MN,A}^{A}+\Gamma_{BN}^{A}\Gamma_{MA}^{B}-\Gamma_{MN}^{B}\Gamma_{BA}^{A}$,
and the 6D Ricci scalar is
$\tilde{R}=\tilde{R}_{MN}\tilde{g}^{MN}$. For our considerations, it is useful to introduce the quantity
\begin{equation}
\tilde{R}_{\rm
4D}=\tilde{g}^{\mu\nu}\big[\Gamma_{\mu\alpha,\nu}^{\alpha}-\Gamma_{\mu\nu,\alpha}^{\alpha}+\Gamma_{\beta\nu}^{\alpha}\Gamma_{\mu\alpha}^{\beta}-\Gamma_{\mu\nu}^{\beta}\Gamma_{\beta\alpha}^{\alpha}\big].
\end{equation}
Since the warp factor is only a function of $r$, we have $\Gamma_{\mu\nu}^{\sigma}
=\frac{1}{2}g^{\sigma\rho}(g_{\mu\rho,\nu}+g_{\rho\nu,\mu}-g_{\mu\nu,\rho})$,
and the 4D Ricci scalar $R_{\rm 4D}$ in Eq.~(\ref{eq:curved}) is thus
related to $\tilde{R}_{\rm 4D}$ by
$R_{\rm 4D}=e^{2\sigma(r)}\tilde{R}_{\rm 4D}$.
In Eq.~(\ref{eq:6D-action}), we can
write
$\sqrt{|\tilde{g}|}\tilde{R}=\sqrt{|\tilde{g}|}(\tilde{g}^{\mu\nu}\tilde{R}_{\mu\nu}+\tilde{g}^{55}\tilde{R}_{55}+\tilde{g}^{66}\tilde{R}_{66})$,
where the first term reads
\begin{eqnarray}
\sqrt{|\tilde{g}|}\tilde{g}^{\mu\nu}\tilde{R}_{\mu\nu} & = & \sqrt{|\tilde{g}|}\tilde{g}^{\mu\nu}\big[\Gamma_{\mu\alpha,\nu}^{\alpha}-\Gamma_{\mu\nu,\alpha}^{\alpha}+\Gamma_{\beta\nu}^{\alpha}\Gamma_{\mu\alpha}^{\beta}-\Gamma_{\mu\nu}^{\beta}\Gamma_{\beta\alpha}^{\alpha}\\
& - & \Gamma_{\mu\nu,5}^{5}+\Gamma_{\beta\nu}^{5}\Gamma_{\mu5}^{\beta}+\Gamma_{5\nu}^{\alpha}\Gamma_{\mu\alpha}^{5}-\Gamma_{\mu\nu}^{5}\Gamma_{5A}^{A}
-
\Gamma_{\mu\nu,6}^{6}+\Gamma_{\beta\nu}^{6}\Gamma_{\mu6}^{\beta}+\Gamma_{6\nu}^{\alpha}\Gamma_{\mu\alpha}^{6}-\Gamma_{\mu\nu}^{6}\Gamma_{6A}^{A}\big].\nonumber\end{eqnarray}
Here, the derivative of the Christoffel symbol with respect to~$r$ (or
$\varphi$) can be written as
\begin{equation}\label{eq:derivative}
-\sqrt{|\tilde{g}|}\tilde{g}^{\mu\nu}\Gamma_{\mu\nu,5}^{5}=-\left[\sqrt{|\tilde{g}|}\tilde{g}^{\mu\nu}\Gamma_{\mu\nu}^{5}\right]_{,5}+\sqrt{|\tilde{g}|}\tilde{g}^{\mu\nu}\Gamma_{A5}^{A}\Gamma_{\mu\nu}^{5}-2\sqrt{|\tilde{g}|}\tilde{g}^{\mu\alpha}\Gamma_{\alpha5}^{\nu}\Gamma_{\mu\nu}^{5},
\end{equation}
where we have used the identities
$(\sqrt{|\tilde{g}|})_{,A}=\sqrt{|\tilde{g}|}\Gamma_{BA}^{B}$ and
${\tilde{g}^{\mu\nu}}_{\,\,\,\,\,\, ,5}=-\tilde{g}^{\mu\alpha}\tilde{g}^{\nu\beta}\tilde{g}_{\alpha\beta,5}=-2\tilde{g}^{\mu\alpha}\Gamma_{\alpha5}^{\nu}$
to obtain respectively the second and the third term on the
right-hand side in Eq.~(\ref{eq:derivative}). We thus have
\begin{equation}
\sqrt{|\tilde{g}|}\tilde{g}^{\mu\nu}\tilde{R}_{\mu\nu}=\sqrt{|\tilde{g}|}\tilde{R}_{\rm
4D}-\left[\sqrt{|\tilde{g}|}\tilde{g}^{\mu\nu}\Gamma_{\mu\nu}^{5}\right]_{,5}-\left[\sqrt{|\tilde{g}|}\tilde{g}^{\mu\nu}\Gamma_{\mu\nu}^{6}\right]_{,6}.
\end{equation}
Similarly, one finds
\begin{subequations}\label{eq:R55R66}
\begin{eqnarray}
\sqrt{|\tilde{g}|}\tilde{g}^{55}\tilde{R}_{55}
& = & \left[\sqrt{|\tilde{g}|}\tilde{g}^{55}\Gamma_{A5}^{A}\right]_{,5}-\left[\sqrt{|\tilde{g}|}\tilde{g}^{55}\Gamma_{55}^{A}\right]_{,A}\nonumber\\
&+&
\sqrt{|\tilde{g}|}\tilde{g}^{55}[\Gamma_{\beta5}^{\alpha}\Gamma_{\alpha5}^{\beta}-\Gamma_{\alpha5}^{\alpha}\Gamma_{\beta5}^{\beta}-2\Gamma_{\alpha5}^{\alpha}\Gamma_{65}^{6}],\label{eq:R55}\\
\sqrt{|\tilde{g}|}\tilde{g}^{66}\tilde{R}_{66}\
& = & \left[\sqrt{|\tilde{g}|}\tilde{g}^{66}\Gamma_{A6}^{A}\right]_{,6}-\left[\sqrt{|\tilde{g}|}\tilde{g}^{66}\Gamma_{66}^{A}\right]_{,A}
+
\sqrt{|\tilde{g}|}\tilde{g}^{66}[\Gamma_{\beta6}^{\alpha}\Gamma_{\alpha6}^{\beta}-\Gamma_{\alpha6}^{\alpha}\Gamma_{\beta6}^{\beta}].
\end{eqnarray}
\end{subequations}
In Eqs.~(\ref{eq:R55R66}), we have
\begin{eqnarray}
&&
\sum_{B=5,6}\Big(
\left[\sqrt{|\tilde{g}|}\tilde{g}^{BB}\Gamma_{AB}^{A}\right]_{,B}-\left[\sqrt{|\tilde{g}|}\tilde{g}^{BB}\Gamma_{BB}^{A}\right]_{,A}
-\left[\sqrt{|\tilde{g}|}\tilde{g}^{\mu\nu}\Gamma_{\mu\nu}^{B}\right]_{,B}\Big)
\label{eq:partialderivatives}\\
&=&\left[\sqrt{|\tilde{g}|}(-\tilde{g}^{\mu\nu}\Gamma_{\mu\nu}^{6}+\tilde{g}^{66}\Gamma_{\alpha6}^{\alpha})\right]_{,6}
+\left[\sqrt{|\tilde{g}|}(-\tilde{g}^{\mu\nu}\Gamma_{\mu\nu}^{5}+\tilde{g}^{55}\Gamma_{\alpha5}^{\alpha}-\tilde{g}^{66}\Gamma_{66}^{5}+\tilde{g}^{55}\Gamma_{65}^{6})\right]_{,5},\nonumber
\end{eqnarray}
and in Eq.~(\ref{eq:partialderivatives}) the last two terms can be written as
\begin{equation}\label{eq:2terms}
\left[\sqrt{|\tilde{g}|}(-\tilde{g}^{66}\Gamma_{66}^{5}+\tilde{g}^{55}\Gamma_{65}^{6})\right]_{,5}=
2\sqrt{|\tilde{g}|}\tilde{g}^{55}\Gamma_{\alpha5}^{\alpha}\Gamma_{65}^{6}
+ 2\sqrt{|\tilde{g}|}\tilde{g}^{55} ( (\Gamma_{65}^{6})^2 +
\Gamma_{65,5}^{6}),
\end{equation}
where we have used the
relation~$\tilde{g}^{66}\Gamma_{66}^{5}=-\tilde{g}^{55}\Gamma_{65}^{6}$.
The first term on the right-hand side in Eq.~(\ref{eq:2terms})
cancels the term
$-2\sqrt{|\tilde{g}|}\tilde{g}^{55}\Gamma_{\alpha5}^{\alpha}\Gamma_{65}^{6}$
in Eq.~(\ref{eq:R55}) while the last term yields the cosmological
term~$\sqrt{|\tilde{g}|}\cdot (-2v^2)$. Putting everything together, the total
action $\mathcal{S}$ in Eq.~(\ref{eq:6D-action}) can be written as
$\mathcal{S}=\mathcal{S}_{\rm 4D}+\mathcal{S}_{\textrm{surface}}+\mathcal{S}_{\textrm{mass}}$,
where
~$\mathcal{S}_{\rm 4D}$, the surface terms~$\mathcal{S}_{\textrm{surface}}$,
and the action $\mathcal{S}_{\textrm{mass}}$ giving rise to the graviton mass terms, respectively read
\begin{subequations}
\begin{eqnarray}
\mathcal{S}_{\rm 4D} & = &
M_{6}^{4}\int\textrm{d}^{6}x\sqrt{|\tilde{g}|}\,(\tilde{R}_{\rm 4D}-2v^2),\label{eq:S-4D}\\
\mathcal{S}_{\textrm{surface}} & = & M_{6}^{4}\int\textrm{d}^{6}x\Big{(}2\left[\sqrt{|\tilde{g}|}\tilde{g}^{55}\Gamma_{\alpha5}^{\alpha}\right]_{,5}
+2\left[\sqrt{|\tilde{g}|}\tilde{g}^{66}\Gamma_{\alpha6}^{\alpha}\right]_{,6}\Big{)},\label{eq:S-surface}\\
\mathcal{S}_{\textrm{mass}} & = &
M_{6}^{4}\int\textrm{d}^{6}x\sqrt{|\tilde{g}|}\Big{(}
\tilde{g}^{55}[\Gamma_{\beta5}^{\alpha}\Gamma_{\alpha5}^{\beta}-\Gamma_{\alpha5}^{\alpha}\Gamma_{\beta5}^{\beta}]+
\tilde{g}^{66}[\Gamma_{\beta6}^{\alpha}\Gamma_{\alpha6}^{\beta}-\Gamma_{\alpha6}^{\alpha}\Gamma_{\beta6}^{\beta}]\Big{)}.\label{eq:S-mass}
\end{eqnarray}
\end{subequations}
Let us first consider~$\mathcal{S}_{\textrm{surface}}$, which is given by
\begin{eqnarray}
\mathcal{S}_{\textrm{surface}} & = & M_{6}^{4}\int\textrm{d}^{4}x\int_{0}^{2\pi}\textrm{d}\varphi\int_{0}^{L}\textrm{d}r\Big{(}\left[\sqrt{|\tilde{g}|}\tilde{g}^{55}\tilde{g}^{\mu\nu}\tilde{g}_{\mu\nu,5}\right]_{,5}+\left[\sqrt{|\tilde{g}|}\tilde{g}^{66}\tilde{g}^{\mu\nu}\tilde{g}_{\mu\nu,6}\right]_{,6}\Big{)}.\end{eqnarray}
The $\varphi$-integral over the second term vanishes for
periodic boundary conditions in $\varphi$-direction and the first
term yields
$\mathcal{S}_{\textrm{surface}}=M_{6}^{4}\int\textrm{d}^{4}x\int_{0}^{2\pi}\textrm{d}\varphi\,[\sqrt{|\tilde{g}|}\tilde{g}^{55}\tilde{g}^{\mu\nu}\tilde{g}_{\mu\nu,5}]_{r=0}^{r=L}$,
which vanishes for suitable boundary conditions at~$r=0,L$.
In a simplified form, $\mathcal{S}_{\rm mass}$ can be written as
\begin{equation}
\mathcal{S}_{\textrm{mass}}=M_{6}^{4}\int\textrm{d}^{6}x\sqrt{|\tilde{g}|}\sum_{c=5,6}\Big[-\frac{1}{4}\tilde{g}^{cc}\tilde{g}_{\mu\nu,c}(\tilde{g}^{\mu\nu}\tilde{g}^{\alpha\beta}-\tilde{g}^{\mu\alpha}\tilde{g}^{\nu\beta})\tilde{g}_{\alpha\beta,c}\Big].\label{eq:S-mass-explicit}\end{equation}
With $\tilde{g}_{\mu\nu}=e^{2\sigma(r)}g_{\mu\nu}$,
$|\tilde{g}|=e^{8\sigma(r)}|g|$, and $R_{\rm 4D}=e^{2\sigma(r)}\tilde{R}_{\rm 4D}$, we then arrive at Eq.~(\ref{eq:splitaction}).

\end{document}